\documentclass[twocolumn,10pt, aps]{revtex4-2}%
\usepackage[paperwidth=210mm,paperheight=297mm,centering,hmargin=2cm,vmargin=2.5cm]{geometry}
\usepackage{amsfonts}
\usepackage{amsmath}
\usepackage{amssymb}
\usepackage{bm}
\usepackage{graphicx}
\usepackage{color}
\usepackage{soul}
\usepackage{epsfig}%
\usepackage[dvipsnames]{xcolor}
  \definecolor{bleu_cite}{RGB}{0,0,255}
\usepackage[colorlinks=true,allcolors = black,citecolor=bleu_cite,  ]{hyperref}
\usepackage{siunitx}

\DeclareSymbolFont{matha}{OML}{txmi}{m}{it}
\DeclareMathSymbol{\varv}{\mathord}{matha}{118}

\def\fd{\textcolor{black}}

\graphicspath{{/home/fd/Nextcloud/Data/Publications/2024/Sub-Radiance/3rd_resub_Nat_Phys_Oct_2025/Figures}}

\begin{document}

\title{Bose-Hubbard simulator with long-range hopping}

\author{Camille Lagoin$^{1}$, Corentin Morin$^{1}$, Kirk  Baldwin$^2$, Loren Pfeiffer$^2$ and Fran\c{c}ois Dubin$^{1,\ddag}$}
\affiliation{$^1$ Université Côte d'Azur, CNRS, CRHEA,Valbonne, France }
\affiliation{$^2$ PRISM, Princeton Institute for the Science and Technology of Materials, Princeton University, Princeton, USA}
\affiliation{$^\ddag$: francois$\_$dubin@icloud.com}

\begin{abstract}
\textbf{Enriching condensed-matter systems with quantum optical phenomena currently drives intense research efforts \cite{douglas2015,chang2018}, particularly to introduce collective quantum correlations \cite{gold_2022,rui_2020,lodahl_2023}. Here we access this paradigm, by confining dipolar excitons in a nanoscopic lattice where long-range hopping, and nearest-neighbour dipolar repulsions, dress the Bose-Hubbard Hamiltonian. Long-range hopping is evidenced by the spontaneous buildup of many-body sub-radiance, signalled by an algebraic slowdown of excitons radiative dissipation \cite{Henriet_2019,chang2018}. In addition, we observe a threshold increase of temporal coherence for dipolar quantum solids only. It suggests that excitons condense in a single sub-radiant state for Mott-like phases. These combine then spatial order and collectively extended coherence, in a single degree of freedom \cite{dubin_2005}. Our study unveils that nanoscopic exciton arrays provide a unique platform to design new frontiers of strongly-correlated lattice models with long-range correlations.}
\end{abstract}

\maketitle

\textbf{Introduction} Semiconductor excitons offer unique opportunities to explore strongly-correlated quantum matter phases. Indeed, excitons, i.e. bound electron-hole pairs, are  massive boson-like quasi-particles, characterized by  Coulomb-mediated inter-particle interactions \cite{Monique_Book}. At the same time, excitons directly couple to the transverse electromagnetic field \cite{hopfield_1958,dubin_2005}. Quantum optical techniques then allow one to efficiently manipulate the exciton wave-function, in the single or few-body regime \cite{lodahl_2023,Gao2012}.

\fd{In the many-body limit, harnessing the usually very strong interactions between excitons constitutes a tedious prerequisite to access predicted collective quantum phases \cite{Monique_Book,Moskalenko_2000}. Recent works have shown that this degree of control} is remarkably achieved in artificial lattices with nanoscopic periods \cite{Joon2023,mak2022,xiong2023,lagoin_2022mott}. \fd{In particular, dipolar excitons of GaAs bilayers realize strongly-interacting bosons confined in electrostatic lattices, and then precisely emulate the Bose-Hubbard Hamiltonian (eBH) extended to interactions between nearest neighbour (NN) lattice sites \cite{Baranov2012,Dutta2015,chanda2025recent}}. Hallmarks of this lattice model have been observed, precisely Mott insulator (MI) phases with one dipolar exciton per lattice site \cite{lagoin_2022,lagoin_2022mott}, and even a long-awaited checkerboard (CB) solid \cite{Baranov2012,Dutta2015,chanda2025recent} stabilized by NN repulsions at half lattice filling \cite{lagoin_2022,lagoin_2024}.

\fd{So far, Bose-Hubbard simulations in GaAs bilayers  have only exploited extended repulsive dipolar interactions between excitons.} However, in electrostatic lattices excitons also realize arrays of optical dipoles with overlapping radiation patterns. Then, it is theoretically expected that collective radiance spontaneously builds up \cite{asenjo2017,Henriet_2019,chang2018}, leading to dissipatively prepared sub-radiant phases \cite{guerin_2016}. These reveal that the interaction between excitons and the transverse electromagnetic field enables a coherent propagation of the optical polarisation in the lattice. Thus, long-range lattice hopping is theoretically introduced \cite{Lehmberg_1970}, corresponding to the recombination of one exciton in a lattice site, yielding a virtual photon reabsorbed at a distant site.

\fd{Here, we demonstrate that dipolar excitons genuinely emulate the Bose-Hubbard Hamiltonian extended by long-range hopping in electrostatic lattices.} Relying on the excitons photoluminescence (PL) dynamics, we show that radiative dissipation is algebraically slowed down in the ultra-low collisional regime. This theoretically expected power-law scaling \cite{Henriet_2019} manifests that \fd{coherent in-plane propagation, combined to extended dipolar repulsions, suppress the conventional exponential dynamics of dissipation \cite{Bouganne2020,Poletti_2012,Moelmer_2020}. At long times, many-body states are then restricted to sub-radiant sub-spaces. For exciton arrays lacking spatial order, we observe that the bandwidth of occupied sub-radiant manifolds is significant, the PL time coherence lying around 15 ps.} By contrast, for spatially ordered dipolar quantum solids, MI and CB, we observe a threshold increase of temporal coherence. It suggests exciton condensation in a single sub-radiant state so that crystalline structure and collective time coherence are strikingly combined.

\textbf{Array of optical dipoles} We use an array of metal gate-electrodes to define a 250 nm period electrostatic lattice in a GaAs bilayer (\fd{Methods}). The lattice confines optically injected dipolar excitons made by electrons and holes lying in a different quantum well. Our device is operated in a blockade regime where two-body on-site dipolar repulsions largely exceed the lattice depth, double occupancy being then prohibited. As illustrated in Fig.1.a, this situation is captured by a Bose-Hubbard Hamiltonian extended to NN interactions  \cite{Baranov2012,Dutta2015}, which reads $H_\mathrm{eBH}=-t\hspace{-.15cm}\sum\limits_{<k,l>}\hspace{-.2cm}B^\dag_kB_l + V/2\sum\limits_{<k,l>}n_k n_l$. Here, $B_k$ is the annihilation operator of one exciton in a site $k$, whose occupation is then  $n_k=B^\dag_kB_k$, while both summations are restricted to NN sites. $V$ denotes the strength of dipolar repulsions between NNs, with a magnitude around 50 $\mu$eV \cite{lagoin_2022,lagoin_2024} greatly exceeding the NN tunnelling strength $t\sim2$ $\mu$eV (Fig.1.a). \fd{Experimentally, we verify that excitons accurately emulate $H_\mathrm{eBH}$ by confirming that MI and CB solids realize the insulating ground-states at $T=330$ mK, for unitary and half lattice fillings respectively (Methods and Extended Data Fig.2).}

Along with inter-particle interactions, excitons directly couple to the transverse electromagnetic field \cite{dubin_2005,Monique_Book,hopfield_1958}, with a radiative dissipation responsible for the PL studied in the following. For our lattice device, Fig.1.b shows that an exciton confined in one site is characterized by an optical radiation pattern that extends over very distant sites. In this regime, theoretical studies have shown that the exciton polarisation propagates across the lattice \cite{asenjo2017,chang2018}. Tracing out the field degree-of-freedom, the global Hamiltonian \fd{reads then $H=H_{eBH}+H_{LR}$ where $H_{LR}=\sum\limits_{k,l}(J_{kl}-i\Gamma_{kl}/2)B^\dag_k B_l$ provides the contribution due to long-range hopping \cite{asenjo2017,Lehmberg_1970}}. The corresponding unitary and dissipative parts, given by $J$ and $\Gamma$, are defined by the real and imaginary projections of the electric-field Green-function respectively. Both are computed by considering a single exciton dipole in an otherwise empty lattice \cite{asenjo2017,Novotny_2012}.\\

\onecolumngrid

\includegraphics[width=\linewidth]{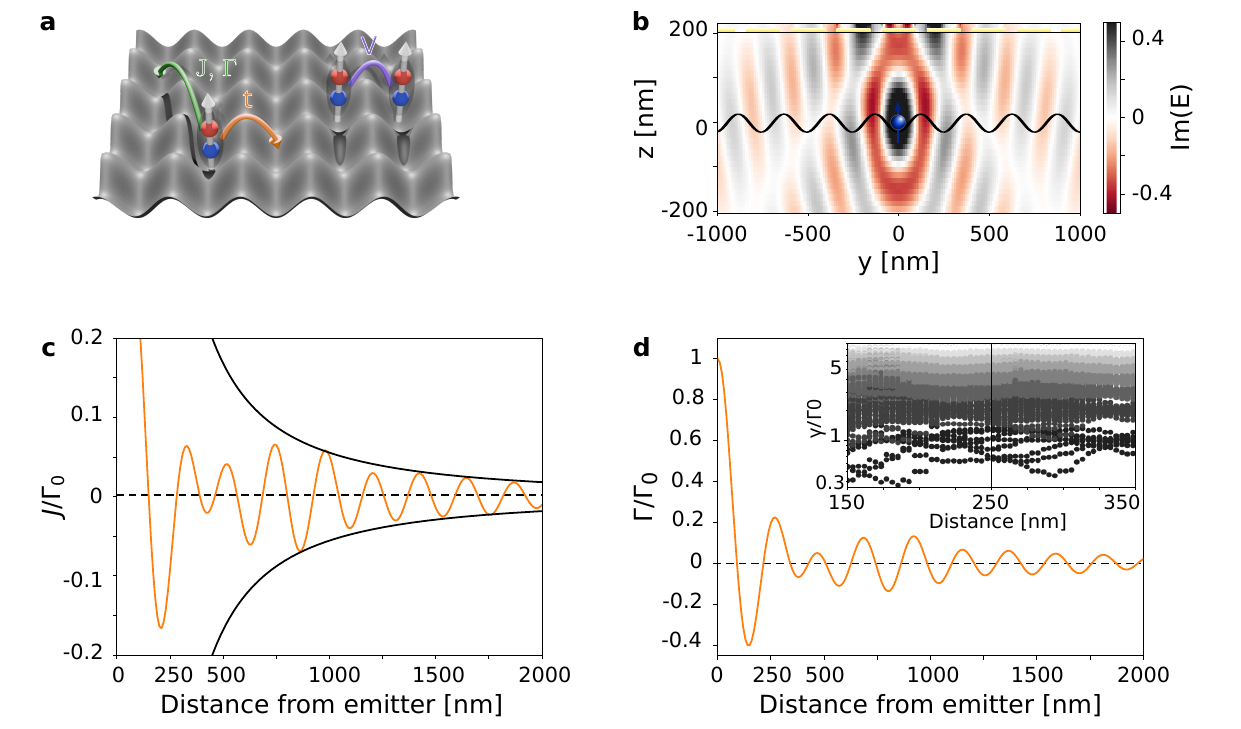}\vspace{.2cm}
\textbf{Fig.1: Bose-Hubbard model with long-range hopping.} \textbf{a} In the occupation blockade regime, the dipolar BH Hamiltonian is governed by the NN tunneling strength $t$, and NN repulsive dipolar interactions with amplitude $V$. Photon-mediated correlations further add a long-range hopping characterized by its real and imaginary parts, $J$ and $\Gamma$ respectively. \textbf{b} Renormalized imaginary part of the electric field (Im(E)) radiated by one exciton dipole confined in one lattice site. \textbf{c-d} Spatial profiles of $J$ and $\Gamma$, taken along one axis of the lattice, normalised by the excitons bare radiative decay rate $\Gamma_0$. The black lines in \textbf{c} define a $1/|r|^{1.6}$ envelope. The inset in \textbf{d}  displays the decay rates of the 2$^9$ states accessible to a 3x3 exciton cluster, emphasizing the region close to our lattice period (vertical line).\\

\twocolumngrid

Figure 1.c quantifies the spatial profile of $J$ along one symmetry axis of the lattice. It shows that $J$ decays non exponentially and instead its envelope exhibits a power-law scaling 1/$|r|^{1.6}$ \fd{at moderate distance $r$}, implying that lattice sites are virtually all coherently correlated. Similarly, Fig.1.d shows that $\Gamma$ takes significant values well above the length scale set by the optical wavelength ($\sim$230 nm in GaAs given the refractive index), revealing that dissipative correlations also extend across all the lattice. To support this conclusion, \fd{we computed the imaginary part $\gamma$ taken by the eigen-values of $H_{LR}$,} for an elementary 3x3 exciton cluster. For our lattice period,  the inset in Fig.1.d shows that manifolds emerge among the $2^9$ accessible states. These are characterized by radiative decay rates increased/decreased up to 5-fold. Hence, we verify that super- and sub-radiant pathways are favoured for our device geometry, \fd{directly confirming the contribution of long-range hopping to the lattice Hamiltonian $H$.}

\textbf{Sub-radiant exciton dynamics} \fd{The dissipative part of long-range hopping, $\Gamma$, is directly accessed by the PL dynamics. Conveniently, our experiments allow us to quantitatively compare the regime where dipolar excitons are confined in the lattice, below unitary filling, to the reference situation where excitons explore instead a homogeneous two-dimensional landscape.} The latter case is for instance accessed at short delays $\Delta t$ after termination of the laser pulse injecting electronic carriers (\fd{Methods}). In this regime where the exciton density is large (around 10$^{10}$ cm$^{-2}$), Fig.2.a illustrates that the PL spectral width is about 900 $\mu$eV. This manifests that a high rate of dipolar collisions screens the lattice potential (around 220 $\mu$eV deep) \cite{ivanov2004,schindler2008}. \fd{The PL signal then decays exponentially with $\Delta t$, like $e^{-\Gamma_0 \Delta t}$ where $\Gamma_0=1/35$ ns$^{-1}$ denotes the bare radiative decay rate of dipolar excitons in our GaAs bilayer (Fig.2.d).} Importantly, the same magnitude for $\Gamma_0$ is measured outside the lattice where excitons explore a flat confinement profile (green points in Fig.2.d). Also, it agrees quantitatively with theoretical expectations for our device geometry \cite{Ivanov_2011}.

\fd{Figure 2.a shows that the PL dynamics is strongly modified when the average exciton density is decreased to around 10$^9$ cm$^{-2}$. This dilute regime is accessed when $\Delta t$ exceeds about 200 ns. The PL spectral width lies then around a few 100 $\mu$eV, which ensures that the lattice confines at most one exciton per site \cite{lagoin_2022mott,lagoin_2022}.  In this situation, Fig.2.b first shows that the PL emission contracts spatially,} from around 30x30 to 10x10 sites. Most strikingly its intensity hardly decays with $\Delta t$, up to the $\mu$s range (Fig.2.a). This considerable increase of the PL radiative decay time suggests that a sub-radiant many-body phase is spontaneously prepared at long times, in the ultra-low collisional regime enforced by the lattice confinement.

\onecolumngrid

\centerline{\includegraphics[width=\textwidth]{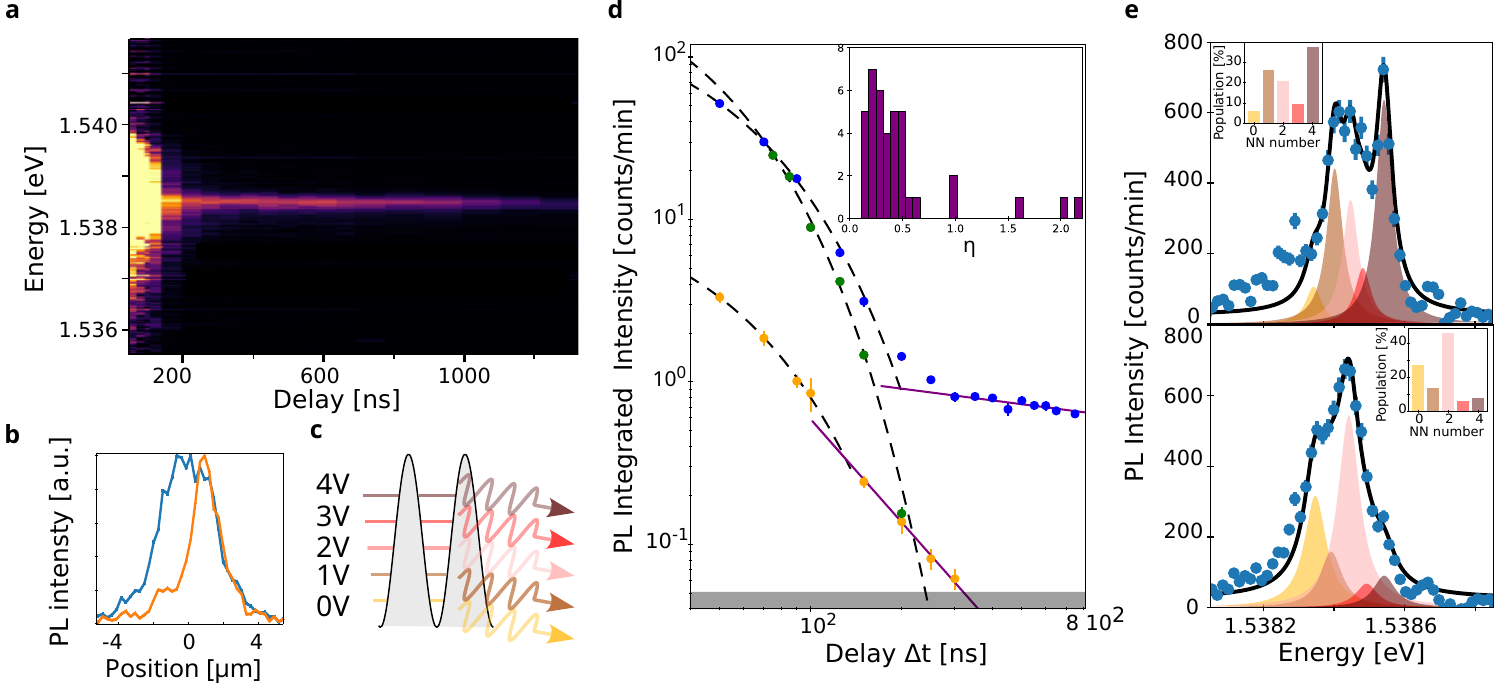}}
\textbf{Fig.2: Sub-radiant PL dynamics.} \textbf{a} Time and spectrally resolved PL measured when the mean power $P$ of the loading laser is set to 20 nW. \textbf{b} Spatial extension of the PL for $\Delta t$= 125 ns (blue) and $\Delta t$= 250 ns (orange). \textbf{c} Accessible energy states in the lattice sites as a function of the number $n$ of NNs. Each level radiates a PL at an energy shifted by $nV$ (wavy arrows). \textbf{d} Dynamics of the PL integrated intensity, at $P=30$ nW (blue) and $P$= 6 nW (orange). Both experiments exhibit the same initial exponential decay with $1/\Gamma_0$= 35 ns time constant (dashed lines), but different algebraic scalings $\Delta t^{-\eta}$ (lines), with $\eta$= 0.2  and 2 at 30 nW and 6 nW respectively. Green points provide the PL decay measured in a region where the top electrode is not structured. The inset shows the occurrence of $\eta$ over 35 experiments.  \textbf{e} PL spectra measured for $\Delta t$= 350 ns (top) and $\Delta t$=550 ns (bottom). Solid lines provide model spectra obtained by adjusting the fraction of sites with 0 to 4 NNs (insets), for $V=50$ $\mu$eV. Experiments were all and performed at 330 mK, for an applied bias (-0.85 V). Error-bars quantify intensity fluctuations, while the gray area in \textbf{d} marks the limit of our PL detection.\\

\twocolumngrid

\fd{In lattices, sub-radiance is theoretically signalled by an algebraic suppression of the radiative decay, from the single to the many-body regimes \cite{chang2018,Henriet_2019}. Our experiments quantitatively obey this power-law scaling, as shown in Fig.2.d highlighting that, for 250 ns $\lesssim\Delta t\lesssim$ 800 ns, the PL dynamics varies like  $\Delta t^{-\eta}$ with $\eta\sim$0.2 (blue). This scaling thus manifests that a photonic channel mediates long-range exciton hopping. This conclusion is supported by Extended Data Fig.1, which compares the PL dynamics for decreasing lattice periods. Hence, we experimentally verify that the algebraic PL decays only buildup when excitons have optical polarisations extending across numerous sites, which requires that the lattice period is below around 350 nm (Methods).}

Figure 2.d further highlights that radiative dissipation is not bound to a single algebraic exponent. Indeed, when the average filling $\bar{N}$ is prepared at about 0.2 and 1, \fd{across the same 8x8 sites and} 250 ns after extinction of the loading laser pulse (blue and orange respectively), $\eta$ decreases from 2 to 0.2. This behaviour suggests that  excitons explore sub-radiant sub-spaces characterized by different densities of states \cite{chang2018,Henriet_2019}. The experiments displayed in Fig.2.d further verify that darkest sub-radiant manifolds, i.e. with lowest $\eta$, are best accessed for largest initial occupations, as theoretically expected \cite{Henriet_2019}. The inset in Fig.2.d illustrates that $\eta$ has in fact a wide range of accessible values, when $\bar{N}$ is varied from 0.1 to 1 between 35 experiments. \fd{Indeed, a rich algebraic dissipative dynamics is robustly observed experimentally (see Supplementary Information for additional details).}

In the lattice, excitons have accessible energies given by their number $n$ of NNs, yielding an energy shift $nV$ (Fig.2.c). This structure is directly captured by the PL spectrum, since the fractions of excitons with 0 to 4 NNs lead to PL lines at different energies (Fig.2.c). We illustrate this property in Fig.2.e, which presents spectral profiles at two different fillings, measured by varying $\Delta t$ for the same experimental conditions. For both situations, where excitons lack spatial order in the lattice, the spectra are quantified solely by adjusting the average fraction of occupied sites with 0 to 4 NN interactions (see insets), setting a spectral width of 75 $\mu$eV for each of their PL emission lines, and $V=50$ $\mu$eV. Hence, we deduce that sub-radiant phases include up to 5 distinguishable modes. However, let us underline that long-range hopping only occurs between lattice sites with the same number of NNs (Fig.2.b). Also, we note that during the dissipative dynamics sub-radiant modes are incoherently coupled by NN tunneling ($t$) and radiative recombinations, both varying $n$ for a given site.

\textbf{Spatial-order and phase coherence} To signal the coherent part of long-range hopping, $J$, spatial interferometry provides a direct approach. However, \fd{Fig.2.b} shows that sub-radiant modes are observed through an around 2 $\mu$m extended region, i.e. too close to our spatial resolution (1.5 $\mu$m) to rely on this technique. Nevertheless, it has been shown that the coherence induced in collective phenomena between excitons is possibly accessed by monitoring the PL degree of temporal coherence \cite{Dang_2020,Prokofev2018}.  \fd{This Fourier transform spectroscopy is performed by measuring the interference contrast $\mathcal{V}$ between the PL emitted at two times separated by a controlled delay $\tau$ (Supplementary Information). The PL time coherence is hence revealed by the decay of $\mathcal{V}(\tau)$, which provides the Fourier transform of the PL spectrum \cite{bell2012}.}

\onecolumngrid
\vspace{.5cm}

\centerline{\includegraphics[width=\linewidth]{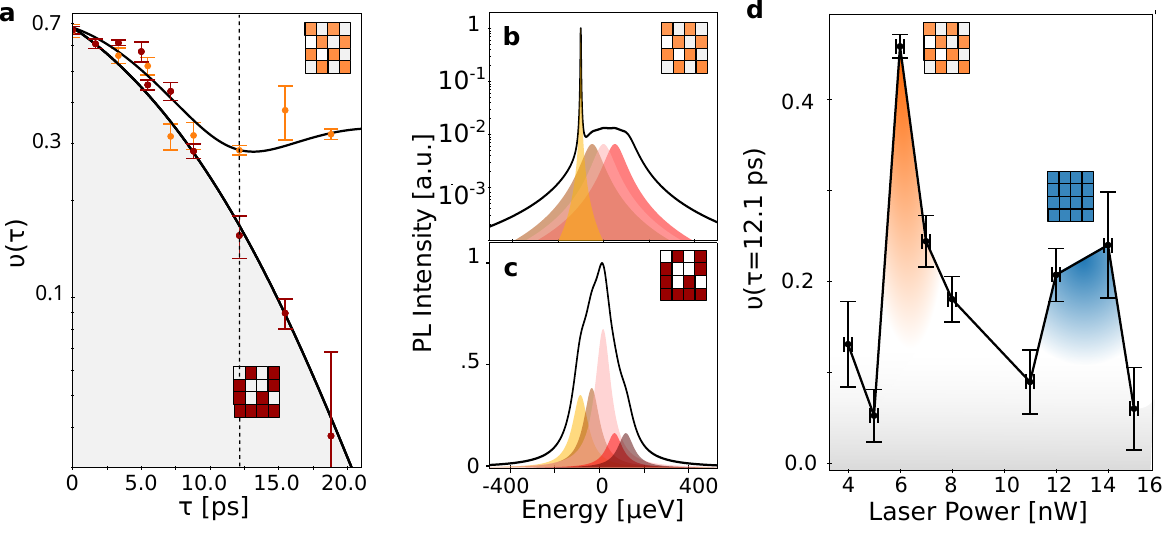}}
\textbf{Fig.3: Time coherence and spatial order.} \textbf{a} $\mathcal{V}(\tau)$ as a function of the optical delay $\tau$, for CB order at half-filling (orange) and for a compressible phase lacking spatial  order around 0.7 filling (red). Lines provide the variations deduced by Fourier transform of the PL spectra shown in \textbf{b} and \textbf{c} for $\bar{N}=$ 1/2 and 0.7 respectively. \textbf{d} $\mathcal{V}$ for  $\tau=12.1$ ps as a function of $\bar{N}$, varied by the laser excitation power $P$. Half and unity lattice fillings are realised around $P$= 6 nW (orange) and 14 nW (blue)  respectively (\fd{Methods}). Measurements were all performed at 330 mK, \fd{at $\Delta t$= 250 ns and $\Delta t$= 400 ns in \textbf{a} and \textbf{d} respectively}, error-bars quantifying our statistical precision.\\

\twocolumngrid

Figure 3.a  displays \fd{$\mathcal{V}(\tau)$} for a phase lacking spatial order realised \fd{across 8x8 sites} at around 0.7 filling (red). Strikingly, the decay of \fd{$\mathcal{V}$} is well captured (solid line) by the Fourier transform  of the adjusted PL spectral profile shown in Fig.3.c. \fd{The latter includes all possible NN interactions, i.e. from 0 to 4NNs, which only have adjusted amplitudes (see coloured shaded areas).} This spectral profile matches the one independently measured in comparable experimental conditions (Fig.2.e). Importantly, in both Fig.2.e and Fig.3.a-c, the spectral width of each sub-radiant contribution, from 0 to 4 NNs, is set to 75 $\mu$eV and $V$= 50 $\mu$eV. For compressible phases, we are then led to conclude that time coherence is limited by the fluctuating distribution of occupied lattice sites. Hence, excitons have a phase diffusing with a characteristic time of 15 ps, \fd{yielding occupied sub-radiant states across an around 75 $\mu$eV bandwidth.}

Figure 3.a also displays in orange the amplitude of \fd{$\mathcal{V}(\tau)$} at $\bar{N}\simeq0.5$. Extended Data Fig.1.c verifies that an incompressible CB solid  is thus realized, and strikingly \fd{$\mathcal{V}$} does not decay for $\tau\gtrsim$7 ps. Instead it remains  constant in our measurement range, revealing that excitons exhibit a strongly extended time-coherence. We only estimate a lower bound of the coherence time, over a few 100 ps. Again, the variation of \fd{$\mathcal{V}(\tau)$} is captured by computing the Fourier transform of an adjusted PL spectrum. A quantitative agreement (line in Fig.3.a) is obtained for the profile displayed in Fig.3.b (in logarithmic scale). It is strikingly governed by a single line, \fd{at the energy for the CB arrangement, i.e. for excitons in the lattice free from NN dipolar repulsions}. This line is orders of magnitude more intense than the contribution corresponding to any other spatial pattern, and with a spectral width reduced to a few $\mu$eV. This behaviour strongly suggests that excitons are condensed in a single sub-radiant state, the PL spectral lineshape actually reproducing the seminally expected signature for Bose-Einstein condensation \cite{Moskalenko_2000},

We anticipate that collective phase-coherence is restricted to incompressible states for which excitons explore a single sub-radiant mode. Otherwise NN tunneling ($t$) inevitably yields random phase fluctuations. To verify this expectation we measured \fd{$\mathcal{V}$} \fd{across the same region comprising 8x8 sites}, for a wide range of lattice fillings prepared by adjusting the power $P$ of the laser loading carriers in the device, while $\tau$ was set to 12.1 ps (see vertical line in Fig.3.a). Figure 3.d signals that \fd{$\mathcal{V}$} is only maximised at $\bar{N}\simeq1/2$ for $P\sim6$ nW (orange), and at $\bar{N}\simeq1$ for $P\sim14$ nW (blue). The exciton compressibility is minimized for both regimes (\fd{Extended Data Fig.1.c}), which attests that CB and Mott solids are realized respectively. Hence, in the lattice quantum insulating phases combine spatial order and collectively extended time coherence.

\textbf{Conclusions} In the ultra-low collisional regime, we have shown that dipolar excitons spontaneously realize sub-radiant many-body phases in electrostatic lattices. Long-range hopping thus dresses the lattice Hamiltonian, yielding an algebraic suppression of dissipation greatly exceeding previous reports with cold atoms \cite{rui_2020,Bloch_2023}. Strikingly, we also observed a threshold increase of excitons temporal coherence, only when spatial order is established. Thus, at unity and half fillings MI and CB quantum insulators exhibit collective time coherence. In the latter case, the lattice translational symmetry is broken spontaneously. This leads us to envision that coherent long-range hopping provides a direct route towards supersolid phases at fractional lattice fillings \cite{Baranov2012,Dutta2015}.

\textbf{Acknowledgments} 

We would like to thank P. Filloux, D. Hrabovsky, S. Suffit for support, and M. Holzmann and G. Pupillo for stimulating discussions. We are also gratefull to J.Bloch, D. Chang, J. Dalibard, T. Grass, M. Lewenstein, H.J. Park and X. Xu for their critical reading of the manuscript. Our research has been financially supported by the French Agency for Research (CE-30 with contracts IXTASE and SIX). The work at Princeton University was funded by the Gordon and Betty Moore Foundation through the EPiQS initiative Grant GBMF4420, and by the National Science Foundation MRSEC Grant DMR 1420541.

\textbf{Author contributions}

K.B and L.P. realised the epitaxial growth of the GaAs heterostructure while F.D. and C.L. designed and nano-fabricated the electrostatic lattice. Experimental works, data analysis, and numerical simulations, were all performed by C.L., C.M., and F.D. who also directed this research.

\textbf{Data availability}

Source data supporting all the conclusions raised in this manuscript are available for download upon reasonable request to FD.

\textbf{Financial interest}

The authors declare no competing financial interest.

\vspace{1cm}

\textbf{METHODS}

\vspace{.5cm}

\textbf{Lattice device}\\

Our device replicates the ones detailed in Refs. \cite{lagoin_2022,lagoin_2024}. It relies on two 8 nm wide GaAs quantum wells separated by a 4 nm Al$_{.3}$Ga$_{.7}$As barrier. On the surface of the 350 nm thick field-effect structure, and \SI{200}{\nano\metre} above the quantum wells, we deposited an array of metal gate electrodes (185x70 nm$^2$ and connected by 30 nm thick wires). The array has a square geometry and 250 nm period. It is polarized at (-0.85) Volt to imprint the electrostatic lattice confining optically injected dipolar excitons. The lattice potential has a theoretical depth around 220 $\mu$eV, and comprises 120x120 sites.

Our experiments rely on a 330 mK micro-PL setup. The lattice device is placed at the focus of a 0.9 numerical aperture microscope objective, mounted on a 3-axis piezo-electric actuator inside a He4/He3 cryostat (Heliox from Oxford Inst.). The objective is used to optically excite the sample, and collect the resulting PL. A 100 ns long laser excitation is precisely applied and repeated at 600 kHz to 900 kHz, with a wavelength at resonance with the direct excitons absorption of the two GaAs quantum wells (around 784 nm). The laser excitation extends accross around 15x15 lattices sites, and dipolar excitons buildup once electrons and holes have tunnelled towards minimum energy states, each located in a distinct quantum well. The average laser excitation power $P$ controls the density of injected electron-hole pairs, i.e. the density of dipolar excitons. These latter radiate a PL around 805 nm, monitored at controlled delays to the termination of the laser illumination, using a 500 mm focal-length spectrometer coupled to an intensified CCD (PI-MAX from Princeton Instruments). In the sub-radiant regime, i.e. at long delays $\Delta t\gtrsim$ 200 ns, we integrate the PL signal in 100-200 ns long time intervals to ensure a sufficient signal-to-noise-ratio, at a repetition rate set to either 600 or 900 kHz. Accumulation times then typically reach minutes, since the PL has a spectrally resolved intensity reduced to a few tens counts/s.

\vspace{.5cm}

\textbf{Lattice filling and exciton compressibility}\\

Our lattice only operates in the dipolar blockade regime, so that its sites have occupations that can not exceed unity. As a result, the maximum PL signal is reached when every site is occupied by one exciton. In the sub-radiant regime, i.e. for delays over 200 ns after the loading pulse, Extended Data Fig.2.a-b illustrates that the PL integrated intensity increases before saturating while $P$ is enhanced, \fd{in the regions studied in Figs.2-3}. Note that the panels a and b correspond to the experiments reported in Fig.2 and Fig.3, performed at 250 ns and at 400 ns delays respectively. In both cases, we deduce the excitation power necessary to saturate the PL, and then we identify the $\bar{N}\sim1$ regime. At lower excitations, the average filling is deduced by computing the corresponding integrated intensity ratio, yielding the values quoted for $\bar{N}$ throughout the manuscript.

The magnitudes extracted for $\bar{N}$ are directly supported by monitoring statistical variations of the PL intensity. Thus, we quantify the exciton compressibility $\kappa$, which is only minimized for insulating phases \cite{lagoin_2022,lagoin_2024,Lagoin_2023}. Extended Data Fig.2.c displays $\kappa$ as a function of $P$ for the experiments shown in Fig.3. Critically, we note that $\kappa$ is minimized for $P$ around 6 nW and 14 nW only. At the same time, from Ex. Data Fig.1.b we deduce that $\bar{N}$ lies around 1/2 and 1 respectively, which confirms that CB and Mott solids are realised.

\vspace{.5cm}

\fd{\textbf{Lattice period and algebraic PL dynamics}}\\

\fd{To unambiguously signal that the algebraic PL decay stems from photon-mediated collective correlations, we studied the PL dynamics as a function of the lattice period. Extended Data Fig.1 compares the PL decay for three devices where the lattice period is set to 800 nm (a), 400 nm (b) and 250 nm (c). The former heterostructure was previously studied in Ref.\cite{lagoin_2022mott}. It has a total thickness equal to 600 nm, the double quantum wells lying 450 nm below the surface. The 400 nm lattice was also studied in Ref.\cite{lagoin_2022mott}, and like the device studied here it relies on a 350 nm thick field-effect structure where the quantum wells are placed 200 nm below the surface \cite{lagoin_2022mott,lagoin_2022}. Moreover, note that for each lattice geometry heterostructures are operated such that the average filling $\bar{N}$ is around unity 250 ns after extinction of the laser excitation \cite{lagoin_2022mott,lagoin_2022}. Accordingly, different bias V$_g$ are applied to the gate electrodes, so that the bare decay rate $\Gamma_0$ varies \cite{Ivanov_2011}.}

\fd{For the 800 nm period lattice Extended Data Fig.1.a shows that $\Gamma_0$= 1/158 ns$^{-1}$ for V$_g$= -2V, in good agreement with theoretical expectations \cite{Ivanov_2011}. Furthermore, we note that the PL dynamics follows a mono-exponential decay up to $\bar{N}\ll1$ for $\Delta t$= 700 ns. For the 400 nm period lattice, Extended Data Fig.1.b shows that for $\Delta t\lesssim$ 350 ns the PL dynamics is exponential, with $\Gamma_0$= 1/214 ns$^{-1}$ for V$_g$= -1.4V, as theoretically expected \cite{Ivanov_2011}. On the other hand, at longer delays we observe a deviation from the exponential scaling, with a slowdown of the PL dynamics like $\Delta t^{-1}$. This behaviour suggests the onset of long-range hopping in a very dilute regime ($\bar{N}\sim0.1$). For the 250 nm period lattice, Extended Data Fig.1.c recalls that the PL is marked by a pronounced algebraic dynamics for 0.1 $\lesssim\bar{N}\lesssim1$.}

\fd{To further confirm that long-range hopping only develops for lattice periods less than around 400 nm, we computed the range accessible to the imaginary part of the eigen-values $\gamma$ of $H_{LR}$, in the one-exciton sub-space. For an infinite two-dimensional array, $\gamma$/$\Gamma_0$ is analytically obtained \cite{asenjo2017}. Extended Data Fig.1.d shows that $\gamma$/$\Gamma_0$ is significantly increased/decreased for lattice periods below around 350 nm, in good agreement with the measurements of the PL dynamics shown in the pannels a-c. Finally, for the 250 nm period lattice device, we recover that radiative decay is increased/decreased by over 5-fold, as discussed in the main text by considering a 3x3 elementary array (inset in Fig.1.d).}

\onecolumngrid
\newpage

\centerline{\includegraphics[width=\linewidth]{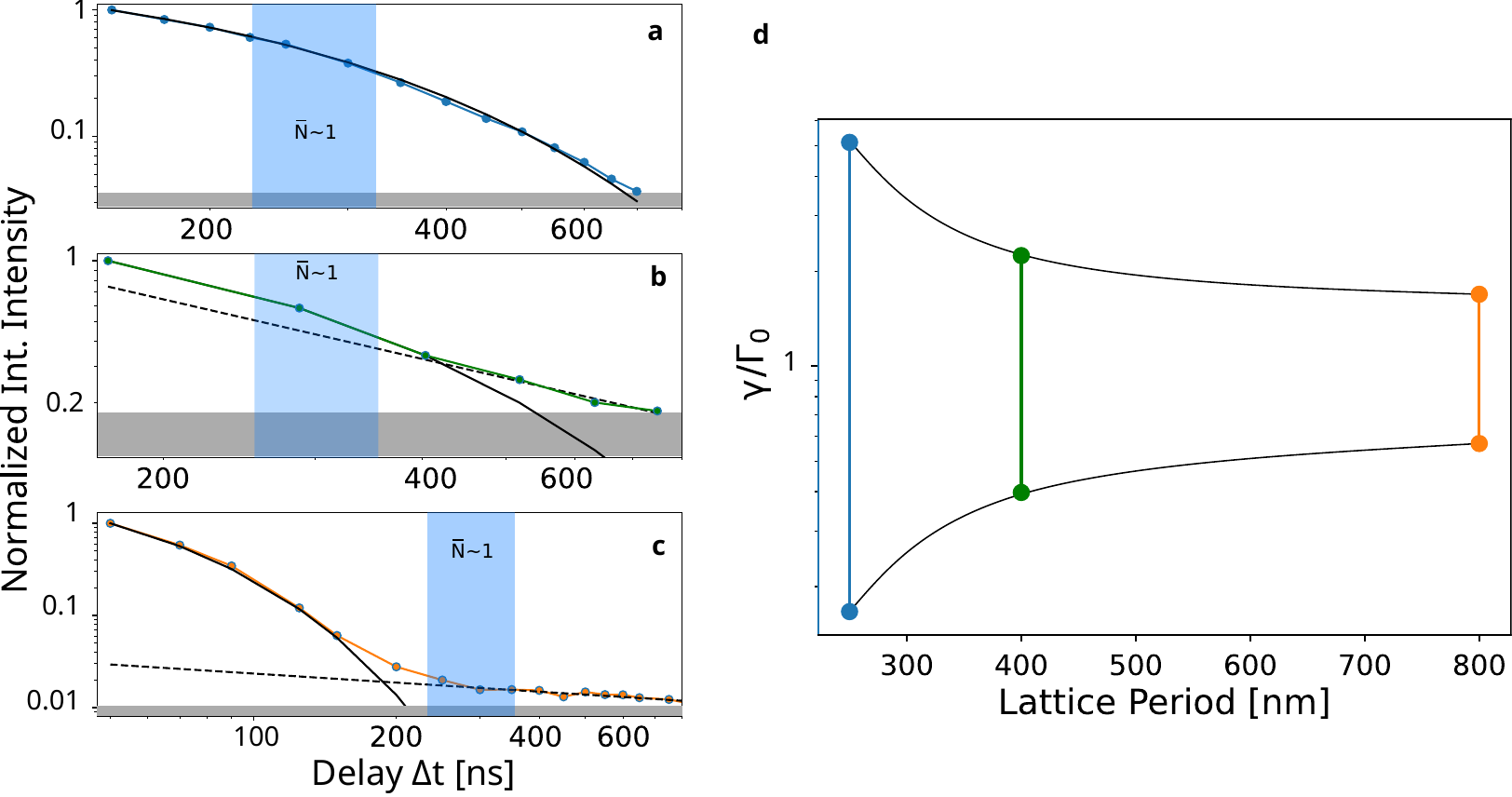}}\vspace{.2cm}
\fd{\textbf{Extended Data Fig.1: Lattice period and algebraic PL decay.} \textbf{a-c} PL decay, in log-log scale, measured in electrostatic lattices with a period equal to 800 nm (\textbf{a}), 400 nm (\textbf{b}) and 250 nm (\textbf{c}) at 330 mK. The 3 devices were studied in a regime where $\bar{N}\sim1$ for $\Delta t\sim$ 250 ns (blue) by suitably tuning the laser excitation power and the gate voltage V$_g$. In \textbf{a} the complete dynamics is quantitatively reproduced by a single exponential decay with a time constant $1/\Gamma_0$= 158 ns (black line). For the 400 nm period lattice (\textbf{b}) the PL decay is initially exponential, with a time constant $1/\Gamma_0$= 214 ns (solid line), and then slows towards an algebraic dynamics $\Delta t^{-1}$ for $\Delta t\gtrsim$ 400 ns (dashed line). Measurements shown in \textbf{c} are obtained with the device discussed in the main text, exhibiting a pronounced algebraic decay at long delays scaling like $\Delta t^{-0.3}$ (dashed line). \textbf{d} Range accessible to the imaginary eigen-values of $H_{LR}$ for each studied lattice period (see colors), computed by assuming an infinite size two-dimensional array \cite{asenjo2017}. The black lines provide a guide for the eye.}

\newpage

\centerline{\includegraphics[width=\linewidth]{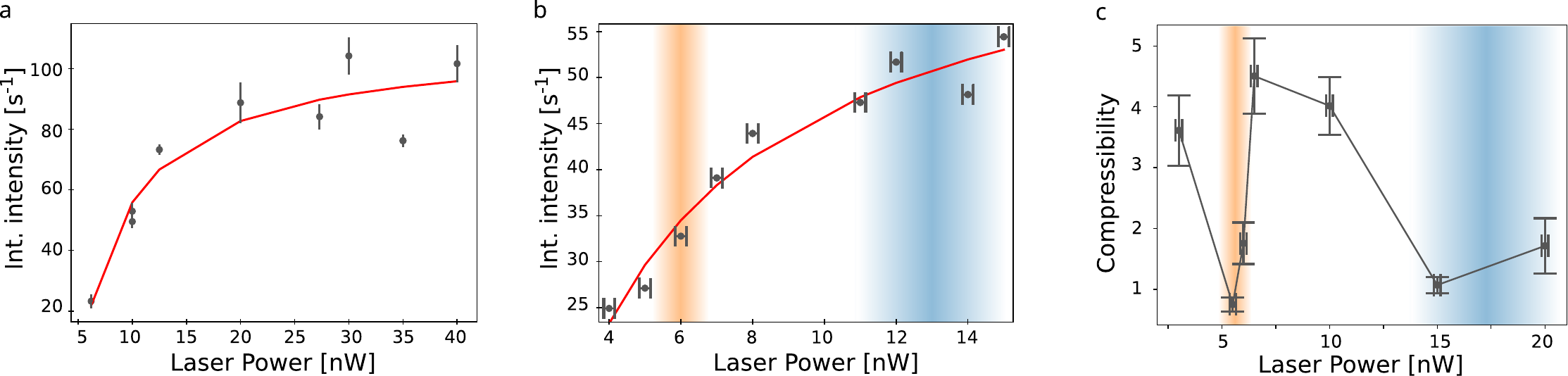}}
\textbf{Extended Data Fig.2: Lattice filling and compressibility.} \textbf{a}-\textbf{b} Integrated intensity of the PL as a function of the average laser excitation power $P$.  The panel \textbf{a} corresponds to the experiments reported in Fig.2. The integrated intensity is measured in a 100 ns long time interval starting 250 ns after extinction of the loading laser pulse repeated at 600 kHz. Panel \textbf{b} corresponds to the measurements shown in Fig.3, acquired 400 ns after extinction of the loading pulse and averaged in a 200 ns long time interval at 900 kHz. \textbf{c} Exciton compressibility normalised to the level of poissonian fluctuations for the experiments shown in \textbf{b}. Experimental data (gray) were all acquired at the lowest 330 mK bath temperature while the red lines in \textbf{a}-\textbf{b} display saturation functions adjusting the measurements. \fd{In (\textbf{b}-\textbf{c}) orange and blue shaded areas underline CB and MI regimes respectively.}

\newpage

\centerline{\Large{\textbf{Supplementary Information for}}}
\centerline{\Large{\textbf{"Bose-Hubbard simulator with long-range hopping"}}}
\vspace{.2cm}
\centerline{\textit{Camille Lagoin$^{1}$, Corentin Morin$^{1}$, Kirk  Baldwin$^2$, Loren Pfeiffer$^2$ and Fran\c{c}ois Dubin$^{1,\ddag}$}}
\centerline{\textit{$^1$ Université Côte d'Azur, CNRS, CRHEA,Valbonne, France}}
\centerline{\textit{$^2$ PRISM, Princeton Institute for the Science and Technology of Materials, Princeton University, Princeton, USA}}
\centerline{\textit{$^\ddag$: francois$\_$dubin@icloud.com}}

\section{Time coherence and Fourier transform spectroscopy}

\fd{In general, time coherence is obtained by studying the decay of the  first-order time-correlation function,  given by
\begin{equation}
g^{(1)}(\tau)=\frac{\langle E^\dag(t)\cdot E(t+\tau)\rangle_t}{\sqrt{\langle |E(t)|^2\rangle_t\langle |E(t+\tau)|^2\rangle_t}}
\end{equation}
where $E$ denotes the PL field and $\langle..\rangle_t$ a statistical average over the time $t$. The intensity spectral distribution $B(\omega)$ allows one to express the average intensity as $\langle |E(t)|^2\rangle_t = \int B(\omega) \,d\omega$. Using the Wiener-Khinchin theorem, the $g^{(1)}$-function then alternatively reads
\begin{equation}
    g^{(1)}(\tau) = \frac{\int B(\omega)exp(i\omega \tau) \,d\omega}{\int B(\omega) \,d\omega}
\end{equation}}

\fd{Experimentally, to access the decay of $g^{(1)}$ we measure the interference contrast of the PL passing through a Mach-Zehnder interferometer (MZI) where an optical delay $\tau$ is induced between the two arms (a and b, see Fig.S1). The output intensity is then possibly expressed as
\begin{equation}
    I_{out} = \langle |E_{out}|^2\rangle_t = \langle |E(t) + \beta E(t+\tau)|^2\rangle_t
\end{equation}
where $E(t)$ and $\beta E(t+\tau)$ denote the PL fields in arm a and b respectively. Here we have explicitly introduced the parameter $\beta$ to account for possible differences between the reflectivity of the two arms, such that $\beta ^2 \le 1$. For simplicity, in the following we consider that fields are real. Thus, we deduce that
\begin{equation}\label{full_expr_Iout}
    I_{out} = \langle |E(t)|^2\rangle_t + \beta^2 \langle |E(t+\tau)|^2\rangle_t + 2 \beta \langle E(t)\cdot E(t+\tau)\rangle_t
\end{equation}}
\fd{so that the last term of Eq(4) is proportional to $g^{(1)}(\tau)$, for which we also note that $\langle |E(t)|^2\rangle_t =\langle |E(t+\tau)|^2\rangle_t = I_0$.}

\fd{Given that the spectral density $B$ can be written as $B(\omega) = 1/2.(B^+(\omega + \omega_0) + B^+(\omega - \omega_0))$ where $B^+(\omega)$ is real and centered around $\omega=0$, we deduce that
\begin{equation}\label{full_expr_Iout_reduced}
    I_{out} = (1+ \beta^2) I_0 + 2 \beta cos(\omega_0 \tau)\int B^+(\omega)exp(i\omega(\tau)) \,d\omega
\end{equation}}

\fd{Usually, the interference contrast $\mathcal{V}$ is defined by
\begin{equation}\label{def_visibility}
    \mathcal{V}(\tau)=\frac{I_{out}^{(cons)}(\tau)-I_{out}^{(des)}(\tau+\delta \tau)}{I_{out}^{(cons)}(\tau)+I_{out}^{(des)}(\tau + \delta \tau)}
\end{equation}
where $I_{out}^{(cons)}$ and $I_{out}^{(des)}$ denote the output intensity for constructive and destructive interference respectively. These correspond to the situations where $\omega_0\tau=0 [2\pi]$ and $\omega_0(\tau+\delta \tau)=\pi [2\pi]$. By considering that the spectral width $ \Delta \omega$ of $B^+(\omega)$ is much smaller than $\omega _ 0$, the variation of $\int_{-\infty}^{\infty}B^+(\omega)exp(i\omega\tau) \,d\omega$ is infinitesimal between times $\tau$ and $\tau + \delta\tau$. We thus obtain
\begin{equation}
    \mathcal{V}(\tau)=\frac{2 \beta \int B^+(\omega)exp(i\omega\tau) \,d\omega}{(1+ \beta^2)I_0}
\end{equation}
which leads to the relation between $\mathcal{V}$ and $g^{(1)}$, namely
\begin{equation}
g^{(1)}(\tau) = cos(\omega_0\tau)\frac{(1+\beta^2)}{2\beta}\mathcal{V}(\tau)
\end{equation}
Hence, we recover that $\mathcal{V}(\tau)$ provides the enveloppe of $g^{(1)}(\tau)$, so that it reveals the PL coherence time.}\\

\centerline{\includegraphics[width=.6\linewidth]{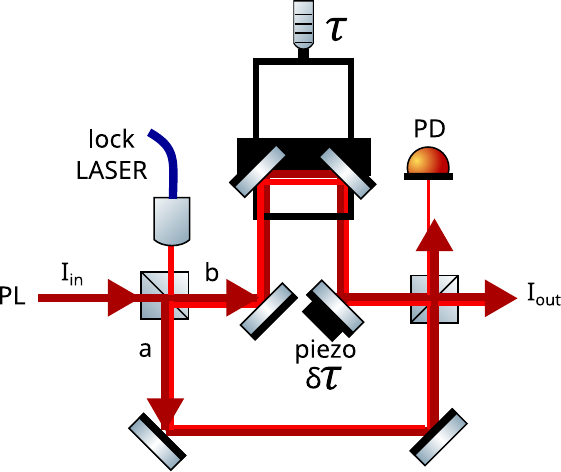}}
\fd{\textbf{Fig.S1: Mach-Zehnder interferometer} Schematic representation of the interferometer used to measure the time coherence of the PL. It splits the  PL (with intensity I$_\mathrm{in}$ at the input) between the two arms a and b, one field being delayed by $\tau$ before the two arms are spatially recombined at the output. Relying on an auxiliary laser beam (light red), the interferometer path length difference is actively stabilized, with 30 nm precision. To this aim, a feedback loop is implemented, acting on the position of a mirror mounted on a piezo-electric stage. Hence, the path length difference is fine tuned by $\delta\tau$, which allows us to compare the situations where the PL interferes constructively and destructively, so that the magnitude of $\mathcal{V}$ is deduced.}\vspace{.3cm}

\subsection{Measurements of the interference contrast}

\centerline{\includegraphics[width=.8\linewidth]{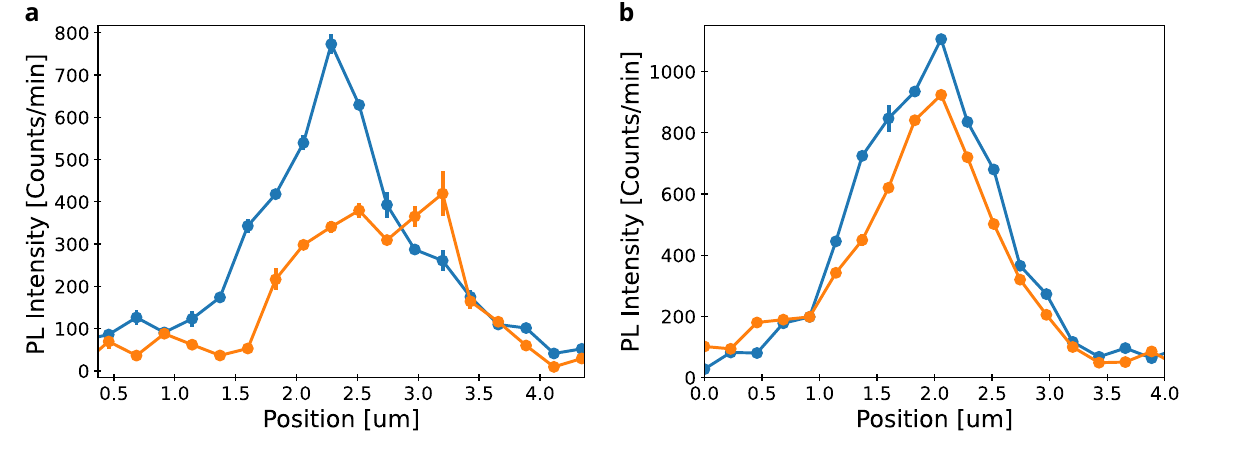}}
\textit{\textbf{Fig.S2: Interference profiles} \textbf{a-b} Interferometer output, spatially resolved, for constructive (blue) and destructive (orange) phase difference. These measurements were realized at 330 mK, 400 ns after extinction of the laser loading pulse, i.e. as in Fig.3.a. For $P$=6 nW (\textbf{a}) and 9.5 nW (\textbf{b}) we set $\bar{N}=1/2$ and $\bar{N}=0.7$ respectively. For a retardation $\tau$=15.5 ps, around 35$\%$ contrast is measured in (\textbf{a}) while the visibility decreases to less than $10\%$ in \textbf{b} (see Fig.3.a).}\\

\fd{Figure S2 presents interferometric results, precisely constructive (blue) and destructive (orange) profiles, spatially resolved along the vertical axis of the lattice. These profiles are obtained by appropriately fixing the relative phase difference $\delta\tau$ between the two arms of the interferometer. Also, let us stress that the interferometer is optically aligned such that the PL equally interferes across its full spatial extension. As shown in Fig.S1, we use a feedback loop to pass from constructive to destructive interference, by acting on the position of one mirror mounted on a piezo-electric transducer. From these two interference profiles we extract the contrast $\mathcal{V}$ for each magnitude of $\tau$. For that, we compute the output intensity, integrated spatially across the profile, and we then deduce $\mathcal{V}$ according to Eq.(6).}

Figure S2.a quantifies the CB phase at $P$=6 nW for $\tau$=15.5 ps. Across over 8x8 sites we deduce that $\mathcal{V}(\tau=15.5\mathrm{ps})$ is around 35$\%$, i.e. about half its value at $\tau$= 0. On the other hand, at $P$=9.5 nW so that $\bar{N}$= 0.7, i.e. when excitons lack spatial order, Fig.S2.b shows that the PL signal varies weakly between constructive and destructive cases, $\mathcal{V}(\tau=15.5\mathrm{ps})$ has dropped to less than 10$\%$, i.e. close to our instrumental precision.

To conclude this section, we compare in Fig.S3 $\mathcal{V}(\tau)$ measured at unity (a) and half (b) fillings. Hence, we directly verify that in both regimes the degree of time coherence does not decay at long retardations $\tau$. It reveals that excitons collectively inherit extended phase coherence for quantum insulators only.\\

\centerline{\includegraphics[width=\linewidth]{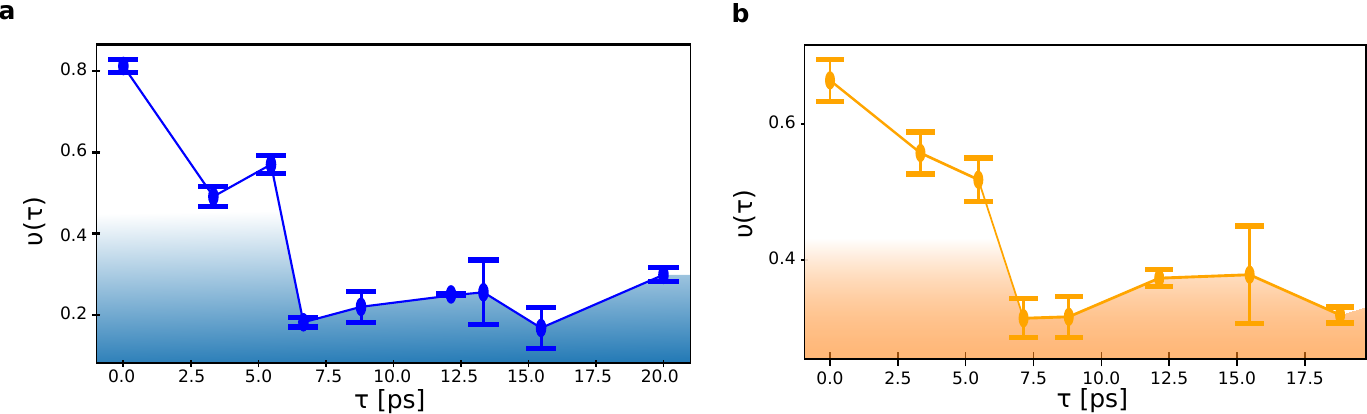}}
\textit{\textbf{Fig.S3: Time coherence of MI and CB solid orders.} \textbf{a} $\mathcal{V}(\tau)$ for MI order at unity-filling. \textbf{b} Identical measurement for CB order at half lattice-filling (orange). Experimental data in \textbf{b} are identical to the ones displayed in Fig.3.a. Measurements were all realised at 330 mK and acquired in a 200 ns long time interval, starting 400 ns after extinction of the loading laser pulse. Error-bars display our statistical precision.}

\section{Bath temperature and sub-radiance}

\centerline{\includegraphics[width=.9\linewidth]{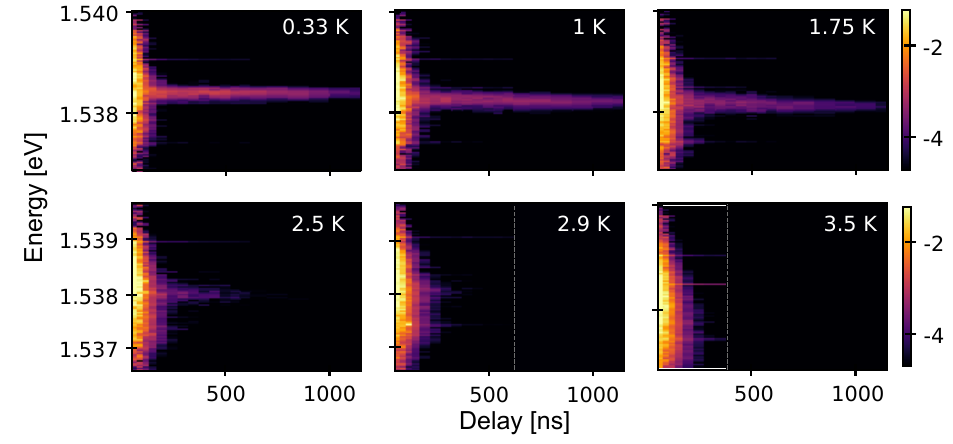}}
\textit{\textbf{Fig.S5: PL decay vs. temperature.} Dynamics of the PL spectrum, displayed in logarithmic scale, as a function of the bath temperature $T$ for $P=20$ nW. In each panel $T$ is specified while the vertical dashed gray lines mark the delay beyond which our signal-to-noise ratio does not allow us to detect any PL.}\\

We monitored the temperature dependence of the PL dynamics. It is displayed in Fig.S5, for $P=20$ nW so that the average filling is set around 0.7, 250 ns after extinction of the loading laser pulse. One directly notes that the long-time  decay, characteristics of sub-radiant phases, varies weakly up to around 2 K, and gradually vanishes at higher temperatures. We interpret this behaviour as the manifestation for a thermal excitation of excitons towards higher-energy states, lying in the continuum above the lattice potential. Indeed, at 2.5 Kelvin the thermal energy compensates the lattice depth. Then sub-radiance is suppressed, as in a flat confinement landscape (Fig.2.d). On the other hand, Fig.S5 highlights that sub-radiance is robust below a few Kelvin, which manifests that accessible states in the lattice are separated by a large energy gap. Indeed, Fig.2.b underlines that the latter is at least $V\sim$50 $\mu$eV. Up to approximately 2 Kelvin, we thus deduce that thermal activation mostly varies the spatial distribution of occupied lattice sites, which does not prevent the buildup of sub-radiance. This behaviour contrasts with quantum order and collective time coherence that are both suppressed above 900-1100 mK, as shown in section III.\\

\section{Temperature and collective time coherence}

Mott and CB solids are protected by energy gaps controlled by the interaction  strengths in the lattice. For our device, bound to the dipolar blockade regime with $V\sim50$ $\mu$eV, we expect that the two quantum insulators remain stable up to around 700 mK and 1 K respectively \cite{lagoin_2022}. Given that extended time coherence is restricted to these two phases only, we studied $\mathcal{V}(\tau)$ as a function of the bath temperature. The corresponding variations are shown in Figures S6.a and S6.b, i.e. at $\bar{N}=1$ and 1/2 respectively. For the latter extended time coherence persists up to approximately 900 mK before abruptly vanishing, reflecting that the CB phase is protected by a gap around 2$V$. On the other hand, Fig.S6.b shows that extended time coherence is maintained up to around 1.1 K at unity filling. This magnitude manifests that Mott order remains robust despite the competition between NN dipolar repulsions and the lattice depth. Indeed, at $\bar{N}\simeq1$ excitons occupy energy states barely confined by the lattice, which fragilises the MI geometry. In fact the Mott phase is reminiscent of a pinned Wigner crystal, as discussed in Ref.\cite{lagoin_2024}.\\

\centerline{\includegraphics[width=\linewidth]{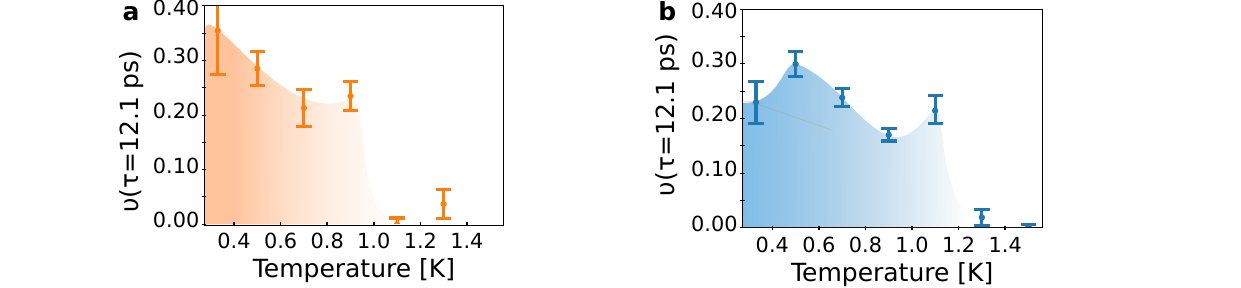}}
\textit{\textbf{Fig.S6: Time-coherence vs. temperature.} $\mathcal{V}$ as a function of the bath temperature, measured for a retardation $\tau$=12.1 ps, for CB (\textbf{a}) and Mott order (\textbf{b}).}\\

\section{Radiative decay of sub-radiant phases}

\centerline{\includegraphics[width=.9\linewidth]{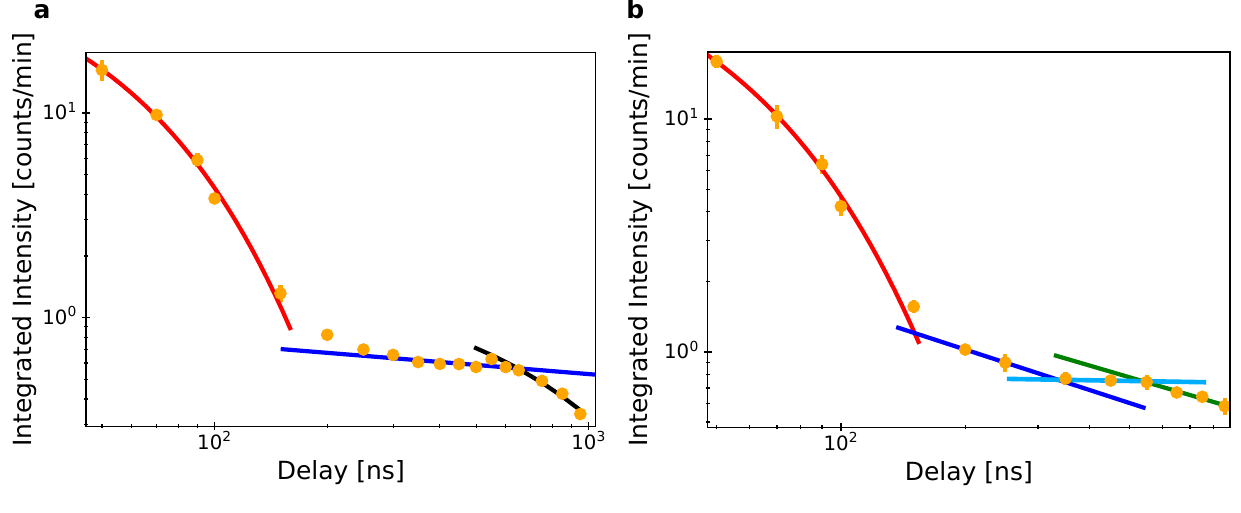}}
\textit{\textbf{Fig.S7: Sub-radiant dynamics} Decay of the PL integrated intensity (orange) measured at $P$=27 nW (\textbf{a}) and 35 nW (\textbf{b}), for a repetition frequency set to 600 kHz. Error-bars display measured intensity fluctuations. Coloured lines highlight the regimes where the decay is exponential (red and black), or algebraic (light/dark blue and green). Experiments were all performed at 330 mK.}\\

In the main text, Fig.2 highlights that sub-radiant modes are signalled by an algebraic time decay, characteristic of the many-body limit. Importantly, the inset in Fig.2.d evidences that algebraic decays are not bound to a single exponent. The latter actually depends strongly on the lattice filling at the earliest delays of the sub-radiant dynamics. In this part, we report measurements showing that, in fact, different algebraic decays are possibly observed successively during the PL decay. For longest delays to the loading laser pulse, we also resolve a transition from algebraic to exponential decay. These behaviours reproduce the theoretically expected dynamics of lattice sub-radiance \cite{Henriet_2019}.

Figure S7.a presents experimental results where one first recovers the intial (free-space) exponential rate of radiative recombinations, with a time constant of 35 ns (red line). It is followed by an algebraic decrease with $\eta=0.15$ (blue line), up to a delay around 650 ns. After, we observe that the radiative dynamics returns to an exponential decay (dark line), but with a time constant of 700 ns. The latter is 20-fold longer than the bare excitons decay time, manifesting that excitons still evolve in a sub-radiant regime.

Figure S7.b displays a second experiment obtained at a slightly higher laser excitation power than in Fig.S7.a. It highlights that two algebraic dynamics successively emerge in the sub-radiant regime (dark blue and green lines), with exponents $\eta=0.5$ and $0.6$ respectively. Strikingly, between these two regimes the PL integrated intensity does not decay at all, during 200 ns where intensity fluctuations are minimized.


\begin{thebibliography}{35}%
\makeatletter
\providecommand \@ifxundefined [1]{%
 \@ifx{#1\undefined}
}%
\providecommand \@ifnum [1]{%
 \ifnum #1\expandafter \@firstoftwo
 \else \expandafter \@secondoftwo
 \fi
}%
\providecommand \@ifx [1]{%
 \ifx #1\expandafter \@firstoftwo
 \else \expandafter \@secondoftwo
 \fi
}%
\providecommand \natexlab [1]{#1}%
\providecommand \enquote  [1]{``#1''}%
\providecommand \bibnamefont  [1]{#1}%
\providecommand \bibfnamefont [1]{#1}%
\providecommand \citenamefont [1]{#1}%
\providecommand \href@noop [0]{\@secondoftwo}%
\providecommand \href [0]{\begingroup \@sanitize@url \@href}%
\providecommand \@href[1]{\@@startlink{#1}\@@href}%
\providecommand \@@href[1]{\endgroup#1\@@endlink}%
\providecommand \@sanitize@url [0]{\catcode `\\12\catcode `\$12\catcode
  `\&12\catcode `\#12\catcode `\^12\catcode `\_12\catcode `\%12\relax}%
\providecommand \@@startlink[1]{}%
\providecommand \@@endlink[0]{}%
\providecommand \url  [0]{\begingroup\@sanitize@url \@url }%
\providecommand \@url [1]{\endgroup\@href {#1}{\urlprefix }}%
\providecommand \urlprefix  [0]{URL }%
\providecommand \Eprint [0]{\href }%
\providecommand \doibase [0]{https://doi.org/}%
\providecommand \selectlanguage [0]{\@gobble}%
\providecommand \bibinfo  [0]{\@secondoftwo}%
\providecommand \bibfield  [0]{\@secondoftwo}%
\providecommand \translation [1]{[#1]}%
\providecommand \BibitemOpen [0]{}%
\providecommand \bibitemStop [0]{}%
\providecommand \bibitemNoStop [0]{.\EOS\space}%
\providecommand \EOS [0]{\spacefactor3000\relax}%
\providecommand \BibitemShut  [1]{\csname bibitem#1\endcsname}%
\let\auto@bib@innerbib\@empty
\bibitem [{\citenamefont {Douglas}\ \emph {et~al.}(2015)\citenamefont
  {Douglas}, \citenamefont {Habibian}, \citenamefont {Hung}, \citenamefont
  {Gorshkov}, \citenamefont {Kimble},\ and\ \citenamefont
  {Chang}}]{douglas2015}%
  \BibitemOpen
  \bibfield  {author} {\bibinfo {author} {\bibfnamefont {J.~S.}\ \bibnamefont
  {Douglas}}, \bibinfo {author} {\bibfnamefont {H.}~\bibnamefont {Habibian}},
  \bibinfo {author} {\bibfnamefont {C.-L.}\ \bibnamefont {Hung}}, \bibinfo
  {author} {\bibfnamefont {A.~V.}\ \bibnamefont {Gorshkov}}, \bibinfo {author}
  {\bibfnamefont {H.~J.}\ \bibnamefont {Kimble}},\ and\ \bibinfo {author}
  {\bibfnamefont {D.~E.}\ \bibnamefont {Chang}},\ }\bibfield  {title} {\bibinfo
  {title} {Quantum many-body models with cold atoms coupled to photonic
  crystals},\ }\href@noop {} {\bibfield  {journal} {\bibinfo  {journal} {Nature
  Photonics}\ }\textbf {\bibinfo {volume} {9}},\ \bibinfo {pages} {326}
  (\bibinfo {year} {2015})}\BibitemShut {NoStop}%
\bibitem [{\citenamefont {Chang}\ \emph {et~al.}(2018)\citenamefont {Chang},
  \citenamefont {Douglas}, \citenamefont {Gonz{\'a}lez-Tudela}, \citenamefont
  {Hung},\ and\ \citenamefont {Kimble}}]{chang2018}%
  \BibitemOpen
  \bibfield  {author} {\bibinfo {author} {\bibfnamefont {D.}~\bibnamefont
  {Chang}}, \bibinfo {author} {\bibfnamefont {J.}~\bibnamefont {Douglas}},
  \bibinfo {author} {\bibfnamefont {A.}~\bibnamefont {Gonz{\'a}lez-Tudela}},
  \bibinfo {author} {\bibfnamefont {C.-L.}\ \bibnamefont {Hung}},\ and\
  \bibinfo {author} {\bibfnamefont {H.}~\bibnamefont {Kimble}},\ }\bibfield
  {title} {\bibinfo {title} {Colloquium: Quantum matter built from nanoscopic
  lattices of atoms and photons},\ }\href@noop {} {\bibfield  {journal}
  {\bibinfo  {journal} {Reviews of Modern Physics}\ }\textbf {\bibinfo {volume}
  {90}},\ \bibinfo {pages} {031002} (\bibinfo {year} {2018})}\BibitemShut
  {NoStop}%
\bibitem [{\citenamefont {Gold}\ \emph {et~al.}(2022)\citenamefont {Gold},
  \citenamefont {Huft}, \citenamefont {Young}, \citenamefont {Safari},
  \citenamefont {Walker}, \citenamefont {Saffman},\ and\ \citenamefont
  {Yavuz}}]{gold_2022}%
  \BibitemOpen
  \bibfield  {author} {\bibinfo {author} {\bibfnamefont {D.}~\bibnamefont
  {Gold}}, \bibinfo {author} {\bibfnamefont {P.}~\bibnamefont {Huft}}, \bibinfo
  {author} {\bibfnamefont {C.}~\bibnamefont {Young}}, \bibinfo {author}
  {\bibfnamefont {A.}~\bibnamefont {Safari}}, \bibinfo {author} {\bibfnamefont
  {T.}~\bibnamefont {Walker}}, \bibinfo {author} {\bibfnamefont
  {M.}~\bibnamefont {Saffman}},\ and\ \bibinfo {author} {\bibfnamefont
  {D.}~\bibnamefont {Yavuz}},\ }\bibfield  {title} {\bibinfo {title} {Spatial
  coherence of light in collective spontaneous emission},\ }\href@noop {}
  {\bibfield  {journal} {\bibinfo  {journal} {PRX Quantum}\ }\textbf {\bibinfo
  {volume} {3}},\ \bibinfo {pages} {010338} (\bibinfo {year}
  {2022})}\BibitemShut {NoStop}%
\bibitem [{\citenamefont {Rui}\ \emph {et~al.}(2020)\citenamefont {Rui},
  \citenamefont {Wei}, \citenamefont {Rubio-Abadal}, \citenamefont {Hollerith},
  \citenamefont {Zeiher}, \citenamefont {Stamper-Kurn}, \citenamefont {Gross},\
  and\ \citenamefont {Bloch}}]{rui_2020}%
  \BibitemOpen
  \bibfield  {author} {\bibinfo {author} {\bibfnamefont {J.}~\bibnamefont
  {Rui}}, \bibinfo {author} {\bibfnamefont {D.}~\bibnamefont {Wei}}, \bibinfo
  {author} {\bibfnamefont {A.}~\bibnamefont {Rubio-Abadal}}, \bibinfo {author}
  {\bibfnamefont {S.}~\bibnamefont {Hollerith}}, \bibinfo {author}
  {\bibfnamefont {J.}~\bibnamefont {Zeiher}}, \bibinfo {author} {\bibfnamefont
  {D.~M.}\ \bibnamefont {Stamper-Kurn}}, \bibinfo {author} {\bibfnamefont
  {C.}~\bibnamefont {Gross}},\ and\ \bibinfo {author} {\bibfnamefont
  {I.}~\bibnamefont {Bloch}},\ }\bibfield  {title} {\bibinfo {title} {A
  subradiant optical mirror formed by a single structured atomic layer},\
  }\href@noop {} {\bibfield  {journal} {\bibinfo  {journal} {Nature}\ }\textbf
  {\bibinfo {volume} {583}},\ \bibinfo {pages} {369} (\bibinfo {year}
  {2020})}\BibitemShut {NoStop}%
\bibitem [{\citenamefont {Tiranov}\ \emph {et~al.}(2023)\citenamefont
  {Tiranov}, \citenamefont {Angelopoulou}, \citenamefont {van Diepen},
  \citenamefont {Schrinski}, \citenamefont {Sandberg}, \citenamefont {Wang},
  \citenamefont {Midolo}, \citenamefont {Scholz}, \citenamefont {Wieck},
  \citenamefont {Ludwig}, \citenamefont {S{\o}rensen},\ and\ \citenamefont
  {Lodahl}}]{lodahl_2023}%
  \BibitemOpen
  \bibfield  {author} {\bibinfo {author} {\bibfnamefont {A.}~\bibnamefont
  {Tiranov}}, \bibinfo {author} {\bibfnamefont {V.}~\bibnamefont
  {Angelopoulou}}, \bibinfo {author} {\bibfnamefont {C.~J.}\ \bibnamefont {van
  Diepen}}, \bibinfo {author} {\bibfnamefont {B.}~\bibnamefont {Schrinski}},
  \bibinfo {author} {\bibfnamefont {O.~A.~D.}\ \bibnamefont {Sandberg}},
  \bibinfo {author} {\bibfnamefont {Y.}~\bibnamefont {Wang}}, \bibinfo {author}
  {\bibfnamefont {L.}~\bibnamefont {Midolo}}, \bibinfo {author} {\bibfnamefont
  {S.}~\bibnamefont {Scholz}}, \bibinfo {author} {\bibfnamefont {A.~D.}\
  \bibnamefont {Wieck}}, \bibinfo {author} {\bibfnamefont {A.}~\bibnamefont
  {Ludwig}}, \bibinfo {author} {\bibfnamefont {A.~S.}\ \bibnamefont
  {S{\o}rensen}},\ and\ \bibinfo {author} {\bibfnamefont {P.}~\bibnamefont
  {Lodahl}},\ }\bibfield  {title} {\bibinfo {title} {Collective super- and
  subradiant dynamics between distant optical quantum emitters},\ }\href@noop
  {} {\bibfield  {journal} {\bibinfo  {journal} {Science}\ }\textbf {\bibinfo
  {volume} {379}},\ \bibinfo {pages} {389} (\bibinfo {year}
  {2023})}\BibitemShut {NoStop}%
\bibitem [{\citenamefont {Henriet}\ \emph {et~al.}(2019)\citenamefont
  {Henriet}, \citenamefont {Douglas}, \citenamefont {Chang},\ and\
  \citenamefont {Albrecht}}]{Henriet_2019}%
  \BibitemOpen
  \bibfield  {author} {\bibinfo {author} {\bibfnamefont {L.}~\bibnamefont
  {Henriet}}, \bibinfo {author} {\bibfnamefont {J.~S.}\ \bibnamefont
  {Douglas}}, \bibinfo {author} {\bibfnamefont {D.~E.}\ \bibnamefont {Chang}},\
  and\ \bibinfo {author} {\bibfnamefont {A.}~\bibnamefont {Albrecht}},\
  }\bibfield  {title} {\bibinfo {title} {Critical open-system dynamics in a
  one-dimensional optical-lattice clock},\ }\href
  {https://doi.org/10.1103/PhysRevA.99.023802} {\bibfield  {journal} {\bibinfo
  {journal} {Physical Review A}\ }\textbf {\bibinfo {volume} {99}},\ \bibinfo
  {pages} {023802} (\bibinfo {year} {2019})}\BibitemShut {NoStop}%
\bibitem [{\citenamefont {Combescot}\ \emph {et~al.}(2022)\citenamefont
  {Combescot}, \citenamefont {Dubin},\ and\ \citenamefont
  {Shiau}}]{dubin_2005}%
  \BibitemOpen
  \bibfield  {author} {\bibinfo {author} {\bibfnamefont {M.}~\bibnamefont
  {Combescot}}, \bibinfo {author} {\bibfnamefont {F.}~\bibnamefont {Dubin}},\
  and\ \bibinfo {author} {\bibfnamefont {S.-Y.}\ \bibnamefont {Shiau}},\
  }\bibfield  {title} {\bibinfo {title} {Signature of electromagnetic quantum
  fluctuations in exciton physics},\ }\href@noop {} {\bibfield  {journal}
  {\bibinfo  {journal} {Europhysics Letters}\ }\textbf {\bibinfo {volume}
  {138}},\ \bibinfo {pages} {36002} (\bibinfo {year} {2022})}\BibitemShut
  {NoStop}%
\bibitem [{\citenamefont {Combescot}\ and\ \citenamefont
  {Shiau}(2015)}]{Monique_Book}%
  \BibitemOpen
  \bibfield  {author} {\bibinfo {author} {\bibfnamefont {M.}~\bibnamefont
  {Combescot}}\ and\ \bibinfo {author} {\bibfnamefont {S.-Y.}\ \bibnamefont
  {Shiau}},\ }\href@noop {} {\emph {\bibinfo {title} {Excitons and Cooper
  Pairs: Two Composite Bosons in Many-Body Physics}}}\ (\bibinfo  {publisher}
  {Oxford University Press},\ \bibinfo {year} {2015})\BibitemShut {NoStop}%
\bibitem [{\citenamefont {Hopfield}(1958)}]{hopfield_1958}%
  \BibitemOpen
  \bibfield  {author} {\bibinfo {author} {\bibfnamefont {J.}~\bibnamefont
  {Hopfield}},\ }\bibfield  {title} {\bibinfo {title} {Theory of the
  contribution of excitons to the complex dielectric constant of crystals},\
  }\href@noop {} {\bibfield  {journal} {\bibinfo  {journal} {Physical Review}\
  }\textbf {\bibinfo {volume} {112}},\ \bibinfo {pages} {1555} (\bibinfo {year}
  {1958})}\BibitemShut {NoStop}%
\bibitem [{\citenamefont {Gao}\ \emph {et~al.}(2012)\citenamefont {Gao},
  \citenamefont {Fallahi}, \citenamefont {Togan}, \citenamefont
  {Miguel-Sanchez},\ and\ \citenamefont {Imamoglu}}]{Gao2012}%
  \BibitemOpen
  \bibfield  {author} {\bibinfo {author} {\bibfnamefont {W.~B.}\ \bibnamefont
  {Gao}}, \bibinfo {author} {\bibfnamefont {P.}~\bibnamefont {Fallahi}},
  \bibinfo {author} {\bibfnamefont {E.}~\bibnamefont {Togan}}, \bibinfo
  {author} {\bibfnamefont {J.}~\bibnamefont {Miguel-Sanchez}},\ and\ \bibinfo
  {author} {\bibfnamefont {A.}~\bibnamefont {Imamoglu}},\ }\bibfield  {title}
  {\bibinfo {title} {Observation of entanglement between a quantum dot spin and
  a single photon},\ }\href {https://doi.org/10.1038/nature11573} {\bibfield
  {journal} {\bibinfo  {journal} {Nature}\ }\textbf {\bibinfo {volume} {491}},\
  \bibinfo {pages} {426} (\bibinfo {year} {2012})}\BibitemShut {NoStop}%
\bibitem [{\citenamefont {Moskalenko}\ and\ \citenamefont
  {Snoke}(2000)}]{Moskalenko_2000}%
  \BibitemOpen
  \bibfield  {author} {\bibinfo {author} {\bibfnamefont {S.}~\bibnamefont
  {Moskalenko}}\ and\ \bibinfo {author} {\bibfnamefont {D.}~\bibnamefont
  {Snoke}},\ }\href@noop {} {\emph {\bibinfo {title} {Bose-Einstein
  condensation of excitons and biexcitons: and coherent nonlinear optics with
  excitons}}}\ (\bibinfo  {publisher} {Cambridge University Press},\ \bibinfo
  {year} {2000})\BibitemShut {NoStop}%
\bibitem [{\citenamefont {Park}\ \emph {et~al.}(2023)\citenamefont {Park},
  \citenamefont {Zhu}, \citenamefont {Wang}, \citenamefont {Wang},
  \citenamefont {Holtzmann}, \citenamefont {Taniguchi}, \citenamefont
  {Watanabe}, \citenamefont {Yan}, \citenamefont {Fu}, \citenamefont {Cao}
  \emph {et~al.}}]{Joon2023}%
  \BibitemOpen
  \bibfield  {author} {\bibinfo {author} {\bibfnamefont {H.}~\bibnamefont
  {Park}}, \bibinfo {author} {\bibfnamefont {J.}~\bibnamefont {Zhu}}, \bibinfo
  {author} {\bibfnamefont {X.}~\bibnamefont {Wang}}, \bibinfo {author}
  {\bibfnamefont {Y.}~\bibnamefont {Wang}}, \bibinfo {author} {\bibfnamefont
  {W.}~\bibnamefont {Holtzmann}}, \bibinfo {author} {\bibfnamefont
  {T.}~\bibnamefont {Taniguchi}}, \bibinfo {author} {\bibfnamefont
  {K.}~\bibnamefont {Watanabe}}, \bibinfo {author} {\bibfnamefont
  {J.}~\bibnamefont {Yan}}, \bibinfo {author} {\bibfnamefont {L.}~\bibnamefont
  {Fu}}, \bibinfo {author} {\bibfnamefont {T.}~\bibnamefont {Cao}}, \emph
  {et~al.},\ }\bibfield  {title} {\bibinfo {title} {Dipole ladders with large
  hubbard interaction in a moir{\'e} exciton lattice},\ }\href@noop {}
  {\bibfield  {journal} {\bibinfo  {journal} {Nature Physics}\ }\textbf
  {\bibinfo {volume} {19}},\ \bibinfo {pages} {1286} (\bibinfo {year}
  {2023})}\BibitemShut {NoStop}%
\bibitem [{\citenamefont {Mak}\ and\ \citenamefont {Shan}(2022)}]{mak2022}%
  \BibitemOpen
  \bibfield  {author} {\bibinfo {author} {\bibfnamefont {K.~F.}\ \bibnamefont
  {Mak}}\ and\ \bibinfo {author} {\bibfnamefont {J.}~\bibnamefont {Shan}},\
  }\bibfield  {title} {\bibinfo {title} {Semiconductor moir{\'e} materials},\
  }\href@noop {} {\bibfield  {journal} {\bibinfo  {journal} {Nature
  Nanotechnology}\ }\textbf {\bibinfo {volume} {17}},\ \bibinfo {pages} {686}
  (\bibinfo {year} {2022})}\BibitemShut {NoStop}%
\bibitem [{\citenamefont {Xiong}\ \emph {et~al.}(2023)\citenamefont {Xiong},
  \citenamefont {Nie}, \citenamefont {Brantly}, \citenamefont {Hays},
  \citenamefont {Sailus}, \citenamefont {Watanabe}, \citenamefont {Taniguchi},
  \citenamefont {Tongay},\ and\ \citenamefont {Jin}}]{xiong2023}%
  \BibitemOpen
  \bibfield  {author} {\bibinfo {author} {\bibfnamefont {R.}~\bibnamefont
  {Xiong}}, \bibinfo {author} {\bibfnamefont {J.~H.}\ \bibnamefont {Nie}},
  \bibinfo {author} {\bibfnamefont {S.~L.}\ \bibnamefont {Brantly}}, \bibinfo
  {author} {\bibfnamefont {P.}~\bibnamefont {Hays}}, \bibinfo {author}
  {\bibfnamefont {R.}~\bibnamefont {Sailus}}, \bibinfo {author} {\bibfnamefont
  {K.}~\bibnamefont {Watanabe}}, \bibinfo {author} {\bibfnamefont
  {T.}~\bibnamefont {Taniguchi}}, \bibinfo {author} {\bibfnamefont
  {S.}~\bibnamefont {Tongay}},\ and\ \bibinfo {author} {\bibfnamefont
  {C.}~\bibnamefont {Jin}},\ }\bibfield  {title} {\bibinfo {title} {Correlated
  insulator of excitons in wse2/ws2 moir{\'e} superlattices},\ }\href@noop {}
  {\bibfield  {journal} {\bibinfo  {journal} {Science}\ }\textbf {\bibinfo
  {volume} {380}},\ \bibinfo {pages} {860} (\bibinfo {year}
  {2023})}\BibitemShut {NoStop}%
\bibitem [{\citenamefont {Lagoin}\ \emph
  {et~al.}(2022{\natexlab{a}})\citenamefont {Lagoin}, \citenamefont {Suffit},
  \citenamefont {Baldwin}, \citenamefont {Pfeiffer},\ and\ \citenamefont
  {Dubin}}]{lagoin_2022mott}%
  \BibitemOpen
  \bibfield  {author} {\bibinfo {author} {\bibfnamefont {C.}~\bibnamefont
  {Lagoin}}, \bibinfo {author} {\bibfnamefont {S.}~\bibnamefont {Suffit}},
  \bibinfo {author} {\bibfnamefont {K.}~\bibnamefont {Baldwin}}, \bibinfo
  {author} {\bibfnamefont {L.}~\bibnamefont {Pfeiffer}},\ and\ \bibinfo
  {author} {\bibfnamefont {F.}~\bibnamefont {Dubin}},\ }\bibfield  {title}
  {\bibinfo {title} {Mott insulator of strongly interacting two-dimensional
  semiconductor excitons},\ }\href@noop {} {\bibfield  {journal} {\bibinfo
  {journal} {Nature Physics}\ }\textbf {\bibinfo {volume} {18}},\ \bibinfo
  {pages} {149} (\bibinfo {year} {2022}{\natexlab{a}})}\BibitemShut {NoStop}%
\bibitem [{\citenamefont {Baranov}\ \emph {et~al.}(2012)\citenamefont
  {Baranov}, \citenamefont {Dalmonte}, \citenamefont {Pupillo},\ and\
  \citenamefont {Zoller}}]{Baranov2012}%
  \BibitemOpen
  \bibfield  {author} {\bibinfo {author} {\bibfnamefont {M.~A.}\ \bibnamefont
  {Baranov}}, \bibinfo {author} {\bibfnamefont {M.}~\bibnamefont {Dalmonte}},
  \bibinfo {author} {\bibfnamefont {G.}~\bibnamefont {Pupillo}},\ and\ \bibinfo
  {author} {\bibfnamefont {P.}~\bibnamefont {Zoller}},\ }\bibfield  {title}
  {\bibinfo {title} {Condensed matter theory of dipolar quantum gases},\
  }\href@noop {} {\bibfield  {journal} {\bibinfo  {journal} {Chem. Rev.}\
  }\textbf {\bibinfo {volume} {112}},\ \bibinfo {pages} {5012} (\bibinfo {year}
  {2012})}\BibitemShut {NoStop}%
\bibitem [{\citenamefont {Dutta}\ \emph {et~al.}(2015)\citenamefont {Dutta},
  \citenamefont {Gajda}, \citenamefont {Hauke}, \citenamefont {Lewenstein},
  \citenamefont {L{\"u}hmann}, \citenamefont {Malomed}, \citenamefont
  {Sowi{\'n}ski},\ and\ \citenamefont {Zakrzewski}}]{Dutta2015}%
  \BibitemOpen
  \bibfield  {author} {\bibinfo {author} {\bibfnamefont {O.}~\bibnamefont
  {Dutta}}, \bibinfo {author} {\bibfnamefont {M.}~\bibnamefont {Gajda}},
  \bibinfo {author} {\bibfnamefont {P.}~\bibnamefont {Hauke}}, \bibinfo
  {author} {\bibfnamefont {M.}~\bibnamefont {Lewenstein}}, \bibinfo {author}
  {\bibfnamefont {D.-S.}\ \bibnamefont {L{\"u}hmann}}, \bibinfo {author}
  {\bibfnamefont {B.~A.}\ \bibnamefont {Malomed}}, \bibinfo {author}
  {\bibfnamefont {T.}~\bibnamefont {Sowi{\'n}ski}},\ and\ \bibinfo {author}
  {\bibfnamefont {J.}~\bibnamefont {Zakrzewski}},\ }\bibfield  {title}
  {\bibinfo {title} {Non-standard hubbard models in optical lattices: a
  review},\ }\href@noop {} {\bibfield  {journal} {\bibinfo  {journal} {Reports
  on Progress in Physics}\ }\textbf {\bibinfo {volume} {78}},\ \bibinfo {pages}
  {066001} (\bibinfo {year} {2015})}\BibitemShut {NoStop}%
\bibitem [{\citenamefont {Chanda}\ \emph {et~al.}(2025)\citenamefont {Chanda},
  \citenamefont {Barbiero}, \citenamefont {Lewenstein}, \citenamefont {Mark},\
  and\ \citenamefont {Zakrzewski}}]{chanda2025recent}%
  \BibitemOpen
  \bibfield  {author} {\bibinfo {author} {\bibfnamefont {T.}~\bibnamefont
  {Chanda}}, \bibinfo {author} {\bibfnamefont {L.}~\bibnamefont {Barbiero}},
  \bibinfo {author} {\bibfnamefont {M.}~\bibnamefont {Lewenstein}}, \bibinfo
  {author} {\bibfnamefont {M.~J.}\ \bibnamefont {Mark}},\ and\ \bibinfo
  {author} {\bibfnamefont {J.}~\bibnamefont {Zakrzewski}},\ }\bibfield  {title}
  {\bibinfo {title} {Recent progress on quantum simulations of non-standard
  bose-hubbard models},\ }\href@noop {} {\bibfield  {journal} {\bibinfo
  {journal} {Reports on Progress in Physics}\ } (\bibinfo {year}
  {2025})}\BibitemShut {NoStop}%
\bibitem [{\citenamefont {Lagoin}\ \emph
  {et~al.}(2022{\natexlab{b}})\citenamefont {Lagoin}, \citenamefont
  {Bhattacharya}, \citenamefont {Grass}, \citenamefont {Chhajlany},
  \citenamefont {Salamon}, \citenamefont {Baldwin}, \citenamefont {Pfeiffer},
  \citenamefont {Lewenstein}, \citenamefont {Holzmann},\ and\ \citenamefont
  {Dubin}}]{lagoin_2022}%
  \BibitemOpen
  \bibfield  {author} {\bibinfo {author} {\bibfnamefont {C.}~\bibnamefont
  {Lagoin}}, \bibinfo {author} {\bibfnamefont {U.}~\bibnamefont
  {Bhattacharya}}, \bibinfo {author} {\bibfnamefont {T.}~\bibnamefont {Grass}},
  \bibinfo {author} {\bibfnamefont {R.}~\bibnamefont {Chhajlany}}, \bibinfo
  {author} {\bibfnamefont {T.}~\bibnamefont {Salamon}}, \bibinfo {author}
  {\bibfnamefont {K.}~\bibnamefont {Baldwin}}, \bibinfo {author} {\bibfnamefont
  {L.}~\bibnamefont {Pfeiffer}}, \bibinfo {author} {\bibfnamefont
  {M.}~\bibnamefont {Lewenstein}}, \bibinfo {author} {\bibfnamefont
  {M.}~\bibnamefont {Holzmann}},\ and\ \bibinfo {author} {\bibfnamefont
  {F.}~\bibnamefont {Dubin}},\ }\bibfield  {title} {\bibinfo {title} {Extended
  bose--hubbard model with dipolar excitons},\ }\href@noop {} {\bibfield
  {journal} {\bibinfo  {journal} {Nature}\ }\textbf {\bibinfo {volume} {609}},\
  \bibinfo {pages} {485} (\bibinfo {year} {2022}{\natexlab{b}})}\BibitemShut
  {NoStop}%
\bibitem [{\citenamefont {Lagoin}\ \emph {et~al.}(2024)\citenamefont {Lagoin},
  \citenamefont {Baldwin}, \citenamefont {Pfeiffer},\ and\ \citenamefont
  {Dubin}}]{lagoin_2024}%
  \BibitemOpen
  \bibfield  {author} {\bibinfo {author} {\bibfnamefont {C.}~\bibnamefont
  {Lagoin}}, \bibinfo {author} {\bibfnamefont {K.}~\bibnamefont {Baldwin}},
  \bibinfo {author} {\bibfnamefont {L.}~\bibnamefont {Pfeiffer}},\ and\
  \bibinfo {author} {\bibfnamefont {F.}~\bibnamefont {Dubin}},\ }\bibfield
  {title} {\bibinfo {title} {Superlattice quantum solid of dipolar excitons},\
  }\href@noop {} {\bibfield  {journal} {\bibinfo  {journal} {Physical Review
  Letters}\ }\textbf {\bibinfo {volume} {132}},\ \bibinfo {pages} {176001}
  (\bibinfo {year} {2024})}\BibitemShut {NoStop}%
\bibitem [{\citenamefont {Asenjo-Garcia}\ \emph {et~al.}(2017)\citenamefont
  {Asenjo-Garcia}, \citenamefont {Moreno-Cardoner}, \citenamefont {Albrecht},
  \citenamefont {Kimble},\ and\ \citenamefont {Chang}}]{asenjo2017}%
  \BibitemOpen
  \bibfield  {author} {\bibinfo {author} {\bibfnamefont {A.}~\bibnamefont
  {Asenjo-Garcia}}, \bibinfo {author} {\bibfnamefont {M.}~\bibnamefont
  {Moreno-Cardoner}}, \bibinfo {author} {\bibfnamefont {A.}~\bibnamefont
  {Albrecht}}, \bibinfo {author} {\bibfnamefont {H.}~\bibnamefont {Kimble}},\
  and\ \bibinfo {author} {\bibfnamefont {D.~E.}\ \bibnamefont {Chang}},\
  }\bibfield  {title} {\bibinfo {title} {Exponential improvement in photon
  storage fidelities using subradiance and ``selective radiance'' in atomic
  arrays},\ }\href@noop {} {\bibfield  {journal} {\bibinfo  {journal} {Physical
  Review X}\ }\textbf {\bibinfo {volume} {7}},\ \bibinfo {pages} {031024}
  (\bibinfo {year} {2017})}\BibitemShut {NoStop}%
\bibitem [{\citenamefont {Guerin}\ \emph {et~al.}()\citenamefont {Guerin},
  \citenamefont {Ara\'ujo},\ and\ \citenamefont {Kaiser}}]{guerin_2016}%
  \BibitemOpen
  \bibfield  {author} {\bibinfo {author} {\bibfnamefont {W.}~\bibnamefont
  {Guerin}}, \bibinfo {author} {\bibfnamefont {M.~O.}\ \bibnamefont
  {Ara\'ujo}},\ and\ \bibinfo {author} {\bibfnamefont {R.}~\bibnamefont
  {Kaiser}},\ }\bibfield  {title} {\bibinfo {title} {Subradiance in a large
  cloud of cold atoms},\ }\href@noop {} {\bibfield  {journal} {\bibinfo
  {journal} {Phys. Rev. Lett.}\ }\textbf {\bibinfo {volume} {116}},\ \bibinfo
  {pages} {083601}}\BibitemShut {NoStop}%
\bibitem [{\citenamefont {Lehmberg}(1970)}]{Lehmberg_1970}%
  \BibitemOpen
  \bibfield  {author} {\bibinfo {author} {\bibfnamefont {R.~H.}\ \bibnamefont
  {Lehmberg}},\ }\bibfield  {title} {\bibinfo {title} {Radiation from an
  {\$}n{\$}-atom system. i. general formalism},\ }\href
  {https://doi.org/10.1103/PhysRevA.2.883} {\bibfield  {journal} {\bibinfo
  {journal} {Physical Review A}\ }\textbf {\bibinfo {volume} {2}},\ \bibinfo
  {pages} {883} (\bibinfo {year} {1970})}\BibitemShut {NoStop}%
\bibitem [{\citenamefont {Bouganne}\ \emph {et~al.}(2020)\citenamefont
  {Bouganne}, \citenamefont {Bosch~Aguilera}, \citenamefont {Ghermaoui},
  \citenamefont {Beugnon},\ and\ \citenamefont {Gerbier}}]{Bouganne2020}%
  \BibitemOpen
  \bibfield  {author} {\bibinfo {author} {\bibfnamefont {R.}~\bibnamefont
  {Bouganne}}, \bibinfo {author} {\bibfnamefont {M.}~\bibnamefont
  {Bosch~Aguilera}}, \bibinfo {author} {\bibfnamefont {A.}~\bibnamefont
  {Ghermaoui}}, \bibinfo {author} {\bibfnamefont {J.}~\bibnamefont {Beugnon}},\
  and\ \bibinfo {author} {\bibfnamefont {F.}~\bibnamefont {Gerbier}},\
  }\bibfield  {title} {\bibinfo {title} {Anomalous decay of coherence in a
  dissipative many-body system},\ }\href
  {https://doi.org/10.1038/s41567-019-0678-2} {\bibfield  {journal} {\bibinfo
  {journal} {Nature Physics}\ }\textbf {\bibinfo {volume} {16}},\ \bibinfo
  {pages} {21} (\bibinfo {year} {2020})}\BibitemShut {NoStop}%
\bibitem [{\citenamefont {Poletti}\ \emph {et~al.}(2012)\citenamefont
  {Poletti}, \citenamefont {Bernier}, \citenamefont {Georges},\ and\
  \citenamefont {Kollath}}]{Poletti_2012}%
  \BibitemOpen
  \bibfield  {author} {\bibinfo {author} {\bibfnamefont {D.}~\bibnamefont
  {Poletti}}, \bibinfo {author} {\bibfnamefont {J.-S.}\ \bibnamefont
  {Bernier}}, \bibinfo {author} {\bibfnamefont {A.}~\bibnamefont {Georges}},\
  and\ \bibinfo {author} {\bibfnamefont {C.}~\bibnamefont {Kollath}},\
  }\bibfield  {title} {\bibinfo {title} {Interaction-induced impeding of
  decoherence and anomalous diffusion},\ }\href
  {https://doi.org/10.1103/PhysRevLett.109.045302} {\bibfield  {journal}
  {\bibinfo  {journal} {Phys. Rev. Lett.}\ }\textbf {\bibinfo {volume} {109}},\
  \bibinfo {pages} {045302} (\bibinfo {year} {2012})}\BibitemShut {NoStop}%
\bibitem [{\citenamefont {Zhang}\ and\ \citenamefont
  {M\o{}lmer}(2019)}]{Moelmer_2020}%
  \BibitemOpen
  \bibfield  {author} {\bibinfo {author} {\bibfnamefont {Y.-X.}\ \bibnamefont
  {Zhang}}\ and\ \bibinfo {author} {\bibfnamefont {K.}~\bibnamefont
  {M\o{}lmer}},\ }\bibfield  {title} {\bibinfo {title} {Theory of subradiant
  states of a one-dimensional two-level atom chain},\ }\href
  {https://doi.org/10.1103/PhysRevLett.122.203605} {\bibfield  {journal}
  {\bibinfo  {journal} {Phys. Rev. Lett.}\ }\textbf {\bibinfo {volume} {122}},\
  \bibinfo {pages} {203605} (\bibinfo {year} {2019})}\BibitemShut {NoStop}%
\bibitem [{\citenamefont {Novotny}\ and\ \citenamefont
  {Hecht}(2012)}]{Novotny_2012}%
  \BibitemOpen
  \bibfield  {author} {\bibinfo {author} {\bibfnamefont {L.}~\bibnamefont
  {Novotny}}\ and\ \bibinfo {author} {\bibfnamefont {B.}~\bibnamefont
  {Hecht}},\ }\href {https://doi.org/DOI: 10.1017/CBO9780511794193} {\emph
  {\bibinfo {title} {Principles of Nano-Optics}}},\ \bibinfo {edition} {2nd}\
  ed.\ (\bibinfo  {publisher} {Cambridge University Press},\ \bibinfo {address}
  {Cambridge},\ \bibinfo {year} {2012})\BibitemShut {NoStop}%
\bibitem [{\citenamefont {Ivanov}(2004)}]{ivanov2004}%
  \BibitemOpen
  \bibfield  {author} {\bibinfo {author} {\bibfnamefont {A.}~\bibnamefont
  {Ivanov}},\ }\bibfield  {title} {\bibinfo {title} {Thermalization and
  photoluminescence dynamics of indirect excitons at low bath temperatures},\
  }\href@noop {} {\bibfield  {journal} {\bibinfo  {journal} {Journal of
  Physics: Condensed Matter}\ }\textbf {\bibinfo {volume} {16}},\ \bibinfo
  {pages} {S3629} (\bibinfo {year} {2004})}\BibitemShut {NoStop}%
\bibitem [{\citenamefont {Schindler}\ and\ \citenamefont
  {Zimmermann}(2008)}]{schindler2008}%
  \BibitemOpen
  \bibfield  {author} {\bibinfo {author} {\bibfnamefont {C.}~\bibnamefont
  {Schindler}}\ and\ \bibinfo {author} {\bibfnamefont {R.}~\bibnamefont
  {Zimmermann}},\ }\bibfield  {title} {\bibinfo {title} {Analysis of the
  exciton-exciton interaction in semiconductor quantum wells},\ }\href@noop {}
  {\bibfield  {journal} {\bibinfo  {journal} {Physical Review B—Condensed
  Matter and Materials Physics}\ }\textbf {\bibinfo {volume} {78}},\ \bibinfo
  {pages} {045313} (\bibinfo {year} {2008})}\BibitemShut {NoStop}%
\bibitem [{\citenamefont {Sivalertporn}\ \emph {et~al.}(2011)\citenamefont
  {Sivalertporn}, \citenamefont {Mouchliadis}, \citenamefont {Ivanov},
  \citenamefont {Philp},\ and\ \citenamefont {Muljarov}}]{Ivanov_2011}%
  \BibitemOpen
  \bibfield  {author} {\bibinfo {author} {\bibfnamefont {K.}~\bibnamefont
  {Sivalertporn}}, \bibinfo {author} {\bibfnamefont {L.}~\bibnamefont
  {Mouchliadis}}, \bibinfo {author} {\bibfnamefont {A.}~\bibnamefont {Ivanov}},
  \bibinfo {author} {\bibfnamefont {R.}~\bibnamefont {Philp}},\ and\ \bibinfo
  {author} {\bibfnamefont {E.}~\bibnamefont {Muljarov}},\ }\bibfield  {title}
  {\bibinfo {title} {Direct and indirect excitons in semiconductor coupled
  quantum wells in an applied electric field},\ }\href
  {https://doi.org/10.1103/PhysRevB.85.045207} {\bibfield  {journal} {\bibinfo
  {journal} {Physical Review B}\ }\textbf {\bibinfo {volume} {85}} (\bibinfo
  {year} {2011})}\BibitemShut {NoStop}%
\bibitem [{\citenamefont {Dang}\ \emph {et~al.}(2020)\citenamefont {Dang},
  \citenamefont {Zamorano}, \citenamefont {Suffit}, \citenamefont {West},
  \citenamefont {Baldwin}, \citenamefont {Pfeiffer}, \citenamefont {Holzmann},\
  and\ \citenamefont {Dubin}}]{Dang_2020}%
  \BibitemOpen
  \bibfield  {author} {\bibinfo {author} {\bibfnamefont {S.}~\bibnamefont
  {Dang}}, \bibinfo {author} {\bibfnamefont {M.}~\bibnamefont {Zamorano}},
  \bibinfo {author} {\bibfnamefont {S.}~\bibnamefont {Suffit}}, \bibinfo
  {author} {\bibfnamefont {K.}~\bibnamefont {West}}, \bibinfo {author}
  {\bibfnamefont {K.}~\bibnamefont {Baldwin}}, \bibinfo {author} {\bibfnamefont
  {L.}~\bibnamefont {Pfeiffer}}, \bibinfo {author} {\bibfnamefont
  {M.}~\bibnamefont {Holzmann}},\ and\ \bibinfo {author} {\bibfnamefont
  {F.}~\bibnamefont {Dubin}},\ }\bibfield  {title} {\bibinfo {title}
  {Observation of algebraic time order for two-dimensional dipolar excitons},\
  }\href {https://doi.org/10.1103/PhysRevResearch.2.032013} {\bibfield
  {journal} {\bibinfo  {journal} {Phys. Rev. Res.}\ }\textbf {\bibinfo {volume}
  {2}},\ \bibinfo {pages} {032013} (\bibinfo {year} {2020})}\BibitemShut
  {NoStop}%
\bibitem [{\citenamefont {Prokof'ev}\ and\ \citenamefont
  {Svistunov}(2018)}]{Prokofev2018}%
  \BibitemOpen
  \bibfield  {author} {\bibinfo {author} {\bibfnamefont {N.~V.}\ \bibnamefont
  {Prokof'ev}}\ and\ \bibinfo {author} {\bibfnamefont {B.~V.}\ \bibnamefont
  {Svistunov}},\ }\bibfield  {title} {\bibinfo {title} {Algebraic time
  crystallization in a two-dimensional superfluid},\ }\href
  {https://doi.org/10.1134/S1063776118110092} {\bibfield  {journal} {\bibinfo
  {journal} {Journal of Experimental and Theoretical Physics}\ }\textbf
  {\bibinfo {volume} {127}},\ \bibinfo {pages} {860} (\bibinfo {year}
  {2018})}\BibitemShut {NoStop}%
\bibitem [{\citenamefont {Bell}(2012)}]{bell2012}%
  \BibitemOpen
  \bibfield  {author} {\bibinfo {author} {\bibfnamefont {R.}~\bibnamefont
  {Bell}},\ }\href@noop {} {\emph {\bibinfo {title} {Introductory Fourier
  transform spectroscopy}}}\ (\bibinfo  {publisher} {Elsevier},\ \bibinfo
  {year} {2012})\BibitemShut {NoStop}%
\bibitem [{\citenamefont {Srakaew}\ \emph {et~al.}(2023)\citenamefont
  {Srakaew}, \citenamefont {Weckesser}, \citenamefont {Hollerith},
  \citenamefont {Wei}, \citenamefont {Adler}, \citenamefont {Bloch},\ and\
  \citenamefont {Zeiher}}]{Bloch_2023}%
  \BibitemOpen
  \bibfield  {author} {\bibinfo {author} {\bibfnamefont {K.}~\bibnamefont
  {Srakaew}}, \bibinfo {author} {\bibfnamefont {P.}~\bibnamefont {Weckesser}},
  \bibinfo {author} {\bibfnamefont {S.}~\bibnamefont {Hollerith}}, \bibinfo
  {author} {\bibfnamefont {D.}~\bibnamefont {Wei}}, \bibinfo {author}
  {\bibfnamefont {D.}~\bibnamefont {Adler}}, \bibinfo {author} {\bibfnamefont
  {I.}~\bibnamefont {Bloch}},\ and\ \bibinfo {author} {\bibfnamefont
  {J.}~\bibnamefont {Zeiher}},\ }\bibfield  {title} {\bibinfo {title} {A
  subwavelength atomic array switched by a single rydberg atom},\ }\href@noop
  {} {\bibfield  {journal} {\bibinfo  {journal} {Nature Physics}\ }\textbf
  {\bibinfo {volume} {19}},\ \bibinfo {pages} {714} (\bibinfo {year}
  {2023})}\BibitemShut {NoStop}%
\bibitem [{\citenamefont {Lagoin}\ \emph {et~al.}(2023)\citenamefont {Lagoin},
  \citenamefont {Suffit}, \citenamefont {Baldwin}, \citenamefont {Pfeiffer},\
  and\ \citenamefont {Dubin}}]{Lagoin_2023}%
  \BibitemOpen
  \bibfield  {author} {\bibinfo {author} {\bibfnamefont {C.}~\bibnamefont
  {Lagoin}}, \bibinfo {author} {\bibfnamefont {S.}~\bibnamefont {Suffit}},
  \bibinfo {author} {\bibfnamefont {K.}~\bibnamefont {Baldwin}}, \bibinfo
  {author} {\bibfnamefont {L.}~\bibnamefont {Pfeiffer}},\ and\ \bibinfo
  {author} {\bibfnamefont {F.}~\bibnamefont {Dubin}},\ }\bibfield  {title}
  {\bibinfo {title} {Dual-density waves with neutral and charged dipolar
  excitons of gaas bilayers},\ }\href
  {https://doi.org/10.1038/s41563-022-01409-9} {\bibfield  {journal} {\bibinfo
  {journal} {Nature Materials}\ }\textbf {\bibinfo {volume} {22}},\ \bibinfo
  {pages} {170} (\bibinfo {year} {2023})}\BibitemShut {NoStop}%
\end{thebibliography}
\end{document}